\documentclass[a4paper,12pt]{article}
\usepackage{psfrag}
\usepackage{graphicx}
\usepackage{graphics}
\usepackage{bm}
\usepackage{color}
\usepackage{amssymb}
\usepackage{amsmath}
\usepackage{float}

\usepackage[T1]{fontenc}
\usepackage[latin1]{inputenc}
\usepackage{amsthm}
\usepackage{braket}
\usepackage{lmodern}
\usepackage{fancyhdr} 
\usepackage{rotating}
\newcommand{\gsim}{\raise.3ex\hbox{$>$\kern-.75em\lower1ex\hbox{$\sim$}}}
\newcommand{\lsim}{\raise.3ex\hbox{$<$\kern-.75em\lower1ex\hbox{$\sim$}}}

\newcommand\ba{{\mathbf{a}}}
\newcommand\bb{{\mathbf{b}}}

\newcommand\bx{{\mathbf{x}}}
\newcommand\bk{{\mathbf{k}}}

\newcommand\bX{{\mathbf{X}}}

\newcommand\be{\begin{equation}}
\newcommand\ee{\end{equation}}
\newcommand\beq{\begin{equation}}
\newcommand\eeq{\end{equation}}
\newcommand\bea{\begin{eqnarray}}
\newcommand\eea{\end{eqnarray}}

\usepackage{vmargin}         
\setmarginsrb{2cm}{2cm}{2cm}{2cm}{0cm}{0cm}{0cm}{1cm}

\title{{\bf Proliferation of sharp kinks on cosmic (super-)string loops with junctions}}

\author{P.~Bin\'etruy\footnote{binetruy@apc.univ-paris7.fr} , A.~Boh\'e\footnote{bohe@apc.univ-paris7.fr} , T.~Hertog\footnote{hertog@apc.univ-paris7.fr} \ 
  and D.A.~Steer\footnote{steer@apc.univ-paris7.fr} \\
{\small {}}\\
{\small {\it  APC\ \footnote{Universit\'e 
Paris-Diderot, CNRS/IN2P3,  CEA/IRFU and Observatoire de Paris} ,10 rue Alice Domon et L\'eonie Duquet,
 75205 Paris Cedex 13, France}}\\
}

\begin{document}
%%%{\small \date{\today}}
\maketitle

\begin{abstract}

Motivated by their effect on the gravitational wave signal emitted by cosmic strings, we study the dynamics of kinks on strings of different tensions meeting at junctions.
The propagation of a kink through a Y-junction leads to the formation of three `daughter' kinks.
Assuming a uniform distribution of the incoming wave vectors at the junction, we find there is a significant region of configuration space in which the sharpness of at least one of the daughter kinks is enhanced relative to the sharpness of the initial kink. For closed loops with junctions we show this leads to an exponential growth in time of very sharp kinks. Using numerical simulations of realistic, evolving cosmic string loops with junctions to calculate the distribution of kink amplitudes as a function of time, we show that loops of this kind typically develop several orders of magnitude of very sharp kinks before the two junctions collide. This collision, or other effects such as gravitational backreaction, may end the proliferation.

\end{abstract}
\maketitle

\section{Introduction}

In recent years there has been a revival in the study of cosmic strings and cosmic superstrings, motivated largely by the realization that cosmic superstrings can form as a by-product of brane inflation (see e.g.~\cite{Polchinski:2004hb,Kibble:2004hq,Davis:2005dd,Myers:1900zz,Copeland:2009ga} for reviews). Of crucial importance is the fact that their observational signatures may provide a unique window on string theory. Hence much work has been dedicated to studying the evolution of cosmic superstring networks (see e.g.~\cite{Tye:2005fn,Hindmarsh:2006qn,Copeland:2005cy,Urrestilla:2007yw,Avgoustidis:2007aa,Avgoustidis:2009ke}) and to determining e.g.~their gravitational wave signatures \cite{DV1,Siemens:2006vk,Siemens:2006yp,Jackson}, which may be detectable by future experiments such as LISA.

The properties of cosmic superstrings differ in at least three ways from those of standard cosmic strings, whose evolution and observational signatures have been studied in depth for over 30 years.  First, a network of cosmic superstrings contains different types of strings, each with a different tension: fundamental F-strings; D1-branes or D-strings; and $(p,q)$-strings which are a bound state of $p$ F-strings and $q$ D-strings.
Second, a network of cosmic superstrings is thought to contain numerous Y-junctions, namely points at which F- and D-strings meet to form the bound state $(p,q)$-string.  Finally, whereas the inter-commutation probability for standard cosmic strings is $P=1$, this is much reduced for cosmic superstrings \cite{Jackson:2004zg}. Indeed, for the collision of two F-strings, $ 10^{-3} \lsim P_F \lsim 1$, whereas for the collision of two D-strings, $0.1 \lsim P_D \lsim 1$.  When strings of different types (such as an F-string and a D-string) collide, they cannot inter-commute due to flux conservation. In certain cases \cite{CKS,CKS2} they form a bound state string with two corresponding Y-junctions.

Earlier studies on the detectability of standard cosmic strings through their gravitational wave emission have recently been partly generalized to cosmic superstrings \cite{DV1,Siemens:2006vk,Siemens:2006yp}. In these papers cosmic superstrings were modeled as usual cosmic strings, but with a reduced inter-commutation probability $P\leqslant 1$. However, until now, the effects of junctions and bound states, as well as the implications of having strings with different tensions, have not been taken into account. 

In this and in a companion paper \cite{Binetruy10b}, we study the effect of junctions on the gravitational wave burst emission from cosmic string networks\footnote{Note that spontaneous symmetry breaking phase transitions can also lead to the formation of cosmic strings with junctions, and hence our analysis also applies to such strings.}.
It is well-known that gravitational radiation from cosmic string loops is dominated by the lowest frequency modes (which are a multiple of the fundamental frequency for loops without junctions) \cite{DePies:2007bm}. Superimposed on the stochastic background of gravitational waves they generate are high frequency bursts emitted at cusps and kinks \cite{Damour}. Kinks in particular radiate as they propagate along a string and through a junction, and when they interact with other kinks \cite{Us}.  Cusps on the other hand are punctual in time, but generate bursts with a higher amplitude. 

For cosmic strings without junctions, cusps provide the dominant contribution to the GW burst signal.
In \cite{Mairii} cusps have been argued to be a generic feature on strings with junctions as well. More recently, however, it has been shown \cite{Ruth} that since cosmic superstrings evolve in a higher dimensional space-time, cusps may be very rare events and furthermore those cusps which are formed are rounded off, hereby significantly reducing their GW burst signal.

For this and other reasons we focus on kinks in this paper. On a loop with no junctions, the number of kinks is constant, fixed by the initial configuration of that loop. However, for a loop containing junctions this is no longer the case. Such loops evolve non-periodically in time and, as we will see, the number of kinks on them increases rapidly: kinks {\it proliferate}. Here we calculate how the number of {\it large amplitude} (that is, very {\it sharp}) kinks proliferates as the loop evolves, since these will turn out to dominate the GW signal from kinks on loops with junctions \cite{Binetruy10b}.

The paper is organized as follows. In Sections \ref{sec:kinkamp} and \ref{sec:prob-dis}, we focus on the interaction between a kink and a single junction. (That is, we do not as yet consider closed loops --- this is done in Sections \ref{section:idealised} and \ref{section:realistic}.) In Section  \ref{sec:kinkamp} we use the dynamical equations of motion for strings with junctions derived in \cite{CKS} to show that when a kink propagates through a Y-junction, it leads to the formation of three `daughter' kinks (one reflected, and two transmitted).  For a specific junction configuration for which the whole evolution can be solved analytically, we show explicitly that the amplitude of the daughter kinks may be {\it larger} than that of the original `parent' kink.\footnote{Kink amplitude, defined in section \ref{sec:prob-dis}, is synonymous to the kink sharpness -- a nomenclature often used elsewhere.} In section  \ref{sec:prob-dis} we generalize this discussion by considering arbitrary junction configurations.  More specifically, we take a uniform distribution of incoming waves at the junction and of incoming kink amplitudes and show that, in a significant region of configuration space, the amplitude of at least one of the daughter kinks is larger than that of the parent kink. 

In Sections \ref{section:idealised} and \ref{section:realistic} we study the evolution of the number of sharp kinks
on closed loops with junctions. In Section \ref{section:idealised}, we consider a simplified model of a loop with junctions that does not take into account the complicated dynamics of the loop itself, and we argue that the number of large amplitude kinks increases exponentially with time. Finally, the dynamics of the loop is taken into account in Section \ref{section:realistic}, where we show that this dynamics generally further enhances the proliferation of sharp kinks. Our conclusions are presented in Section \ref{conc}.

\section{Propagation of a kink through a junction}
\label{sec:kinkamp}

In this section, following \cite{CKS}, we first review the equations of motion for three semi-infinite Nambu-Goto strings  of tensions $\mu_1$, $\mu_2$ and $\mu_3$ which meet at a Y-junction. A consequence of these equations is that the propagation of a kink through a Y-junction results in the production of three `daughter' kinks; a reflected kink as well as two transmitted kinks. We also define kink amplitude (or sharpness), and analytically study -- in the simplest case of an initially static junction -- the amplitude of the daughter kinks, showing that in some cases these can be amplified relative to the incoming kink.

\subsection{Description of the system}

We work in flat spacetime with signature $(- + + +)$, and use the standard conformal-temporal gauge
so that each string is described by its spatial coordinates $\textbf{x} _{j}(\sigma,t)$, where $t$ coincides with Lorentz time and the subscript $j$ labels the different strings. The gauge constraints can then be written as (with $'$ and $ \dot{ }$ standing for derivatives with respect to  $ \sigma$ and $t$ respectively)
\begin{eqnarray}
\textbf{x}'  _{j}\cdot \dot{\textbf{x} } _{j}& = & 0 \\
\textbf{x}  _{j}  ^{\prime 2} + \dot{\textbf{x} } _{j} ^{2} & = & 1.
\label{conf-gauge}
\end{eqnarray}
The action describing the combined system of three semi-infinite strings meeting at a junction has been analysed in \cite{CKS}.
Away from the junction the wave-like equation of motion for each string has solution
\begin{equation}
\label{basic}
\textbf{x} _{j}(\sigma,t)=\frac{1}{2}\big(\textbf{a}_{j}(u)+\textbf{b}_{j}(v)\big)
\end{equation}
where
\be
u=\sigma+t, \qquad v=\sigma-t,
\ee
and where $\textbf{a} _{j}^{\prime 2}=\textbf{b}_{j} ^{\prime 2}=1$ in order to satisfy the gauge constraints. Each string is bounded by the junction located at $\bX(t)=\bx_i(s_i(t),t)$. One can therefore let
$\sigma$ take values in the interval  $]-\infty,s _{j}(t)]$.

As explained in \cite{CKS}, the initial conditions for $\dot{\mathbf{x}}_j$ and $\mathbf{x}'_j$ at $t=0$  determine $\mathbf{a}_j(u)$ and $\mathbf{b}_j(v)$ for $u\leqslant s_j(0)$ and $v \leqslant s_j(0)$. A set of coupled differential equations describing the physics at the junction then enables one to determine the evolution of $s_j(t)$ as well as the outgoing waves $\mathbf{a}'_j(s_j(t)+t)$ in terms of the incoming waves $\mathbf{b}'_j(s_j(t)-t)$ (which are determined by the initial conditions since $\dot{s}_j\leqslant1$).
In particular, let
\bea
\nu_1 &=& \mu_2 + \mu_3 - \mu_1 \geqslant 0
\label{nudef}
\\
M_1&=& \mu_1^2-(\mu_2-\mu_3)^2 =\nu_2 \nu_3 \geqslant 0
\nonumber
\\
c_1(t)&=&\mathbf{b}'_2(s_2(t)-t)\cdot\mathbf{b}'_3(s_3(t)-t),
\label{prodscal}
\eea
and circular permutations, as well as
\be
\mu = \mu_1+\mu_2+\mu_3.
\ee
Then the equations of motion imply energy conservation at the junction
\be
\mu_1 \dot{s}_1 + \mu_2 \dot{s}_2 + \mu_3 \dot{s}_3 =0.
\label{energy}
\ee
One also has 
\begin{equation}
\label{spoint}
\frac{\mu_j(1-\dot{s}_j)}{\mu}=\frac{M_j(1-c_j)}{M_1(1-c_1)+M_2(1-c_2)+M_3(1-c_3)}
\end{equation}
and
\begin{equation}
\label{aprime}
\mathbf{a}'_j\big(s_j(t)+t\big)=
\frac{1}{1+\dot{s}_j} \left[(1-\dot{s}_j) \mathbf{b}'_j\big(s_j(t)-t\big)-
\frac{2}{\mu}\sum_{k=1}^3 \mu_k(1-\dot{s}_k)\mathbf{b}'_k\big(s_k(t)-t\big) \right]
\end{equation}
which determines the outgoing wave on string $j$ in terms of the inward moving waves.  The last term in 
\eqref{aprime} is proportional to the velocity of the junction:
\be
\dot{\bX} = -\frac{1}{\mu}\sum_{k=1}^3 \mu_k(1-\dot{s}_k)\mathbf{b}'_k\big(s_k(t)-t\big) .
\label{X}
\ee

From \eqref{aprime} it follows that if \emph{one of the strings} has a kink propagating towards the junction, i.e.~one of the functions $\mathbf{b}'_j(v)$ has a discontinuity, then all $\mathbf{a}'_j(u)$ acquire a discontinuity when the kink reaches the junction. The presence of the junction therefore increases the number of kinks in the system from 1 to 3. We refer to the three newly formed kinks as `daughter kinks'.

Furthermore, eq.~(\ref{aprime}) also implies that for essentially {\it all} initial 
conditions fixed at $t=0$ say, the subsequent evolution always generates kinks.
Indeed,  from  Eq.~(\ref{aprime}) and for $t=0^+$, $\ba_j'(s_j(0)^+)$ on string $j$ is determined by the weighted sum of {\it all $\bb'_k(s_k(0)^-)$}.  The latter are fixed by the arbitrary initial conditions and independent of the $\ba'_k(s_k(0)^-)$. Their sum therefore yields a vector $\ba'_j(s_j(0)^+)$ which is generally different from $\ba'_j(s_j(0)^-)$. Therefore $\ba'_j$ will be discontinuous\footnote{
The exception is when,
 given the $\bb'_j(s_j(0)^-)$,
the $\ba'_j(s_j(0)^-)$ are chosen to be given by the RHS of eq.~(\ref{aprime}). 
A particular case of this are totally static initial conditions ($\dot{\bx}_j=0=\dot{\bX}=\dot{s}_j$) considered in \cite{Bevis:2009az}.} 
 at $u=s_j(0)$, leading to a kink $t=0^+$. Hence the presence of junctions essentially implies the existence of kinks, which therefore will not need to be introduced by hand in our simulations below.

\subsection{Amplitude and transmission coefficients}

Kinks on cosmic strings are sources of gravitational wave bursts: a kink emits bursts throughout its propagation on the string \cite{Damour}, and also when it  encounters another kink or when it crosses a junction \cite{Us}. However, the GW signal emanating from kinks on a network of strings is determined not only by their number, but also by their amplitudes. The kink amplitude can be defined as follows.

Consider a kink moving towards a junction ({\it i.e.}~a discontinuity of $\bb'_j(v)$ at $v_*$). The amplitude of all GW bursts associated with kinks is proportional to the components of  \cite{Damour,Us} 
\begin{equation}
\label{facteuramplitude}
\Bigg(\frac{\mathbf{b}_j^{\prime}(v_* ^{+})}{1 -  \mathbf{n}\cdot\mathbf{b}_j^{\prime}(v_*^{+})} -  \frac{\mathbf{b}^{\prime}_j(v_*^{-})}{1 -  \mathbf{n}\cdot\mathbf{b}_j^{\prime}(v_*^{-})}\Bigg).
\end{equation}
Both denominators in \eqref{facteuramplitude} are generically comparable and of order one, because the direction of emission $\mathbf{n}$ is a priori uncorrelated with $\mathbf{b}'_j(v_*^+)$ and 
$\mathbf{b}'_j(v_*^-)$, and therefore (\ref{facteuramplitude}) is of order $\| \mathbf{b}'_j(v_*^+)-\mathbf{b}'_j(v_*^-)\|$.  We define the {\it kink amplitude} or {\it sharpness} by
\begin{equation}
\label{defamplitudekink}
A[{\bb'_j}] = \frac{1}{2} \| \mathbf{b}'_j(v_*^+)-\mathbf{b}'_j(v_*^-)\| 
=|\sin(\theta/2)|
\ee
where $\theta$ is the angle between $\mathbf{b}'_j(v_*^{\pm})$. The factor of 1/2 is a normalisation factor so that $0 \leq A[{\bb'_j}]  \leq 1$. We adopt an analogous definition of the sharpness of an outward moving kink on string $j$ characterized by a discontinuity in $\mathbf{a}'_j(u)$.

At first sight one might expect the daughter kinks to have smaller amplitudes than the incoming kink. However, this is not the case. In particular energy conservation at the junction, eq.~(\ref{energy}), does not constrain the amplitude of the transmitted kinks\footnote{The energy along the string is proportional to $\sigma$. Hence no energy is associated to the kink itself which is described by a pointlike discontinuity.}. Indeed the derivatives in (\ref{energy}) suffer a discontinuity when a kink hits the junction, and these therefore undergo a sudden jump.

It will also be useful to define the following {\it transmission coefficients} $C_j$. For instance, consider a kink that propagates towards the junction on string $1$ ({\it i.e.}~$\mathbf{b}'_1(v)$ has a discontinuity at $v_{1*}$) and reaches the junction at $t_*$,  with $v_{1*}=s_1(t_*)-t_*$. Then each function $\mathbf{a}'_j(u)$ acquires a discontinuity at $u_{j*}=s_j(t_*)+t_*$. We define
\begin{equation}
\label{deftransmissioncoefficients}
C_j = \frac{A[\ba'_j]}{A[{\bb'_1}]},
\end{equation}
where the amplitudes are given by eq.~(\ref{defamplitudekink}).

\subsection{Example: static junction}
\label{example-static}

To conclude this section, we illustrate the production of daughter kinks with an example that can be worked out analytically. 

The strings of tension $\mu_1$ and $\mu_2=\mu_3$ are taken to be in the $(x,y)$ plane at all times, and the initial configuration considered is shown in the left-hand panel of Figure 1. String $1$ lies along the $x$-axis, and strings $2$ and $3$ subtend an angle $\psi$ w.r.t the $y$-axis,  chosen such that the junction is {\it initially static}.
There are two kinks on string 1, both of which propagate towards the junction (they are discontinuities in $\bb_1'$). Once both kinks have propagated through the junction, the angles between the strings at the junction is again the same as initially and hence the junction is {\it again static}. However, all three strings now have outward moving kinks on them as shown in the right hand panel of Figure 1. The amplitude of the outgoing kinks can be calculated as follows.

Let
\bea
\bk_1&=&(-1,0)\; 
\nonumber 
\\
\bk_{1}^{\theta}&=&-(\cos \theta,\sin \theta)\; 
\nonumber
\\
\bk_2&=&(\epsilon, -\delta) \;  \nonumber
\\
\bk_3&=&(\epsilon,\delta)\; 
\eea
where $\theta\in[-\pi,\pi]$ is a free parameter and we have defined
\be
0 \leq \epsilon \equiv \frac{\mu_1}{2\mu_2}\leq 1, \qquad \delta \equiv \sqrt{1-\epsilon^2}.
\ee
Hence the angle $\psi = \arctan(\epsilon/\delta)$.
Then the initial condition corresponding to Fig. 1 is given by

\begin{minipage}[t]{0.38\textwidth}
\begin{equation*}
\left\{
\begin{array}{lll}
\mathbf{a}'_1(u) &= &\bk_1 \qquad \text{if } u\leqslant 0\\
\mathbf{a}'_2(u) &= &\bk_2 \qquad \text{if } u\leqslant 0\\
\mathbf{a}'_3(u) &= &\bk_3 \qquad \text{if } u\leqslant 0
\end{array} \right.\\
\end{equation*}
\end{minipage}
\begin{minipage}[t]{0.58\textwidth}
\begin{equation*}
\left\{
\begin{array}{lll}
\mathbf{b}'_1(v) = \left|
\begin{array}{ll}
\bk_1 &  \quad \text{if }  -L \leqslant v \leqslant 0 \\
\bk_1^\theta & \quad \text{if } -L-\ell \leqslant v \leqslant -L\\
\bk_1& \quad \text{if }  v \leqslant -L-\ell\\
\end{array} \right.\\
\mathbf{b}'_2(v) =\bk_2  \qquad \text{if } v\leqslant0\\
\mathbf{b}'_3(v) =\bk_3   \qquad \text{if } v\leqslant0
\end{array}
\right.
\end{equation*}
\end{minipage}\\
%\vspace{0.5cm}
%l
From (\ref{basic}) it now follows that for string 1 and for $ v \in [-L-\ell,-L]$,
 \bea
 \bx'_1&=&\cos(\theta/2)\big(\cos(\theta/2),\sin(\theta/2) \big) \; ,
 \nonumber
 \\
 \dot{\bx}_1&=& \sin(\theta/2)\big(\sin(\theta/2),-\cos(\theta/2) \big) \; .
 \eea
Thus the physical angle to the $x$-axis made by the segment of string 1 between the two kinks is $\theta_{\rm geo}=\theta/2$. It moves with velocity $\sin(\theta/2)$ towards the junction.

\begin{figure}[H]
\begin{minipage}[t]{0.48\textwidth}
   \centering
   \includegraphics[scale=0.25]{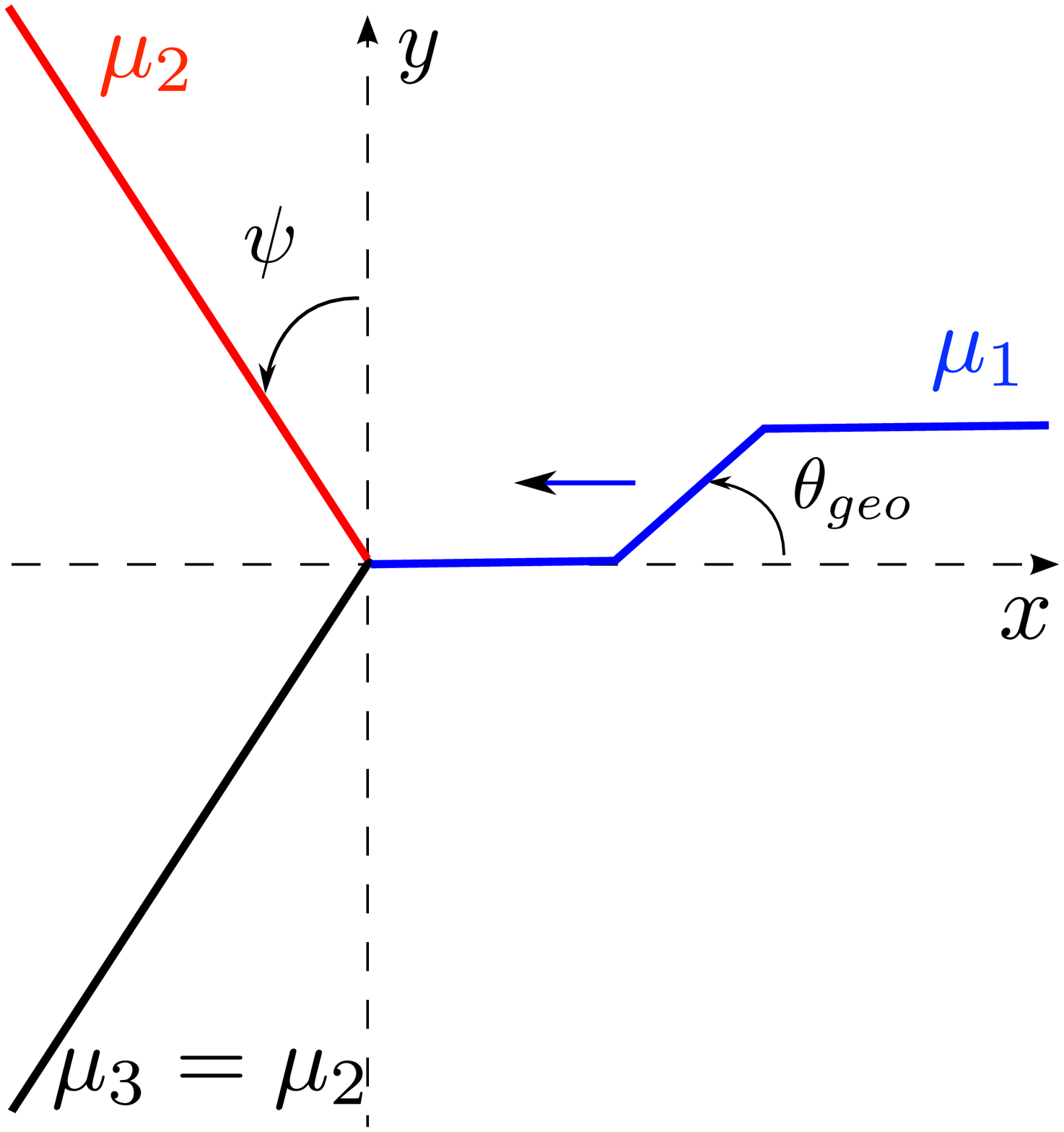} 
\end{minipage}
\begin{minipage}[t]{0.48\textwidth}
   \centering
   \includegraphics[scale=0.25]{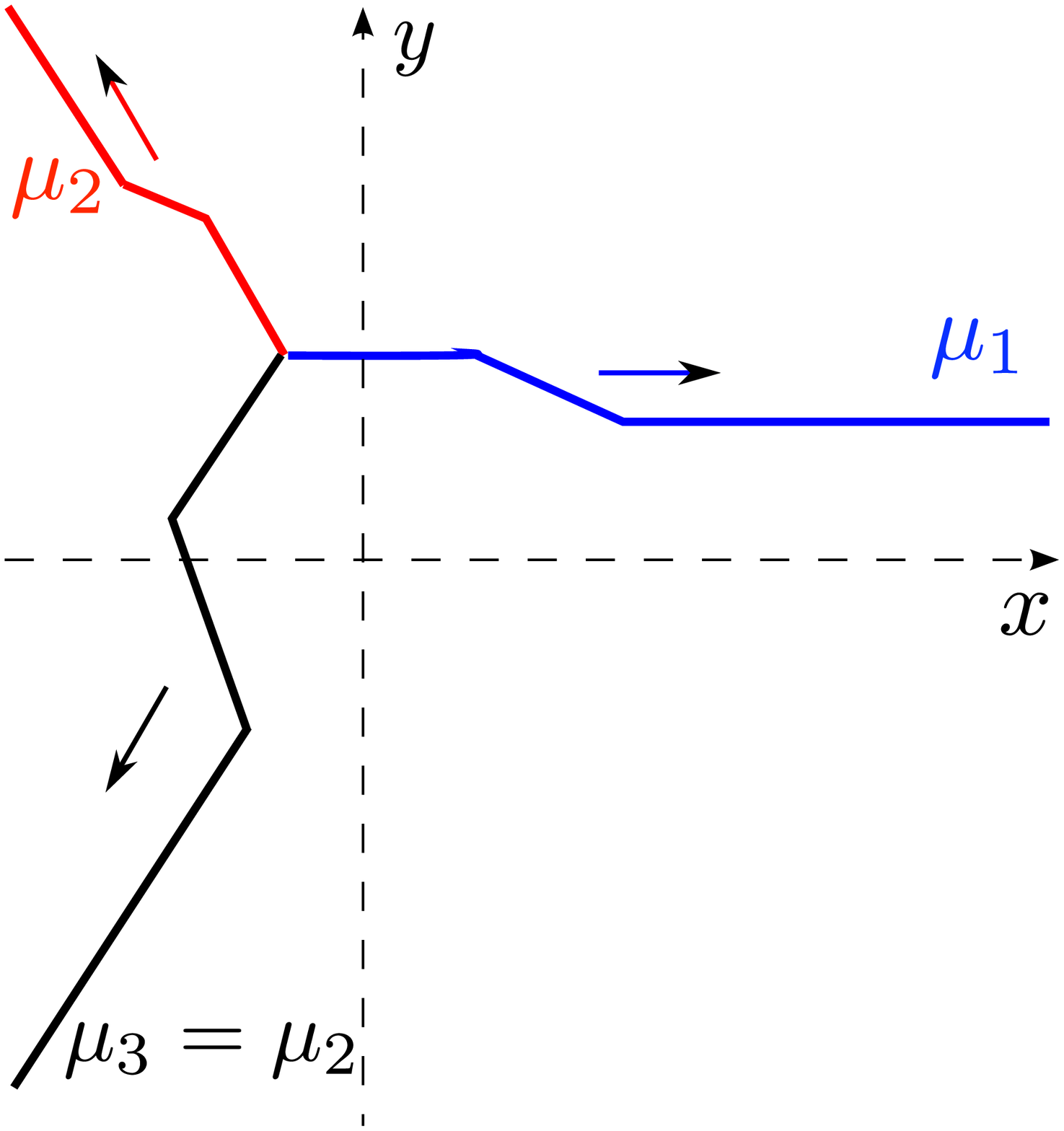} 
\end{minipage}
\caption{Propagation of two kinks through an initially static junction. Initial (left) and final (right) configuration of the system.}
   \label{examplefig}
\end{figure}

From the equations of motion \eqref{prodscal}--\eqref{aprime} one can analytically determine the evolution of the system, since the $\mathbf{b}'_j$ are piecewise constant functions. The evolution consists of three distinct phases:

\begin{itemize}
\item $t\in[0,L]$. 
\end{itemize}

The solutions of $c_j$ in (\ref{prodscal}), substituted into (\ref{spoint}), yield $\dot{s}_j=0$
on each string. Thus (\ref{X}) implies the junction is indeed static: $\dot{\bX}=0$ \cite{CKS}.
The outgoing waves are determined by (\ref{aprime}): $\mathbf{a}'_j(u)=\mathbf{k}_j$ for $u\in [0,L]$. Hence during this phase, the segment between the two kinks propagates towards the junction with velocity $|\sin(\theta/2)|$ whilst the rest of the system remains static. This phase ends when $t=L$, because then $s_1(t=L)-L=v=-L$ so the scalar products $c_j(t)$ (which depend on $\mathbf{b}'_1(s_1(t)-t)$) change.

\begin{itemize}
\item $t\in[L,t_*]$.
\end{itemize}

During this phase the segment between the two kinks crosses the junction and $\bb_1'=\bk_1^\theta$. The time $t_*$ is the solution to $s_1(t_*)-t_*=-L-\ell$. Beyond this time, $\bb_1'$ changes back to $\bk_1$ again.
From \eqref{spoint} we find that the $\dot{s}_j$ are {\it constant} and equal to
\begin{eqnarray}
\dot{s}_1 & = & {\frac {\epsilon \left( \cos \theta  -1 \right)}{\epsilon\,\cos \theta +(1+ \epsilon +\epsilon^2)}} \; , \\
\dot{s}_2 & = & 
{\frac {(1-\cos\theta ){\epsilon}^{2}-\delta\,\sin \theta (1+
\epsilon)}{\epsilon\,\cos \theta +(1+ \epsilon +\epsilon^2)}}\; , 
\label{sdot2}
\\
\dot{s}_3 & = & {\frac {(1-\cos\theta ){\epsilon}^{2}+\delta\,\sin \theta (1+
\epsilon)}{\epsilon\,\cos \theta +(1+ \epsilon +\epsilon^2)}}=-(\dot{s}_1 + \dot{s}_2)\; .
\end{eqnarray}
Thus $s_j(t)=\dot{s}_j(t-L)$, so that $t_*=L+\frac{\ell}{1-\dot{s}_1}$.
The functions $\ba_j'$ can then be obtained from \eqref{aprime}; for instance
\be
\mathbf{a}'_1(u)=
\frac{1}{1+\epsilon^2+2\epsilon \cos\theta} \big(-(1+\epsilon^2)\cos\theta -2\epsilon , \delta^2 \sin\theta \big) 
\qquad u\in\left[L,L+\frac{1+\dot{s}_1}{1-\dot{s}_1}\right] \, .
\ee
Finally eq.~(\ref{X}) determines the constant velocity of the junction during this time interval;
\be
\dot{\bX} = \frac{1}{1+\epsilon + \epsilon^2+\epsilon \cos\theta} \big(\epsilon (\cos\theta-1), (1+\epsilon \sin\theta) \big) \, .
\ee

\begin{itemize}
\item $t>t_*$. 
\end{itemize}

The configuration at the junction is now exactly the same as during the first phase. That is, $\dot{s}_j=0$ on each string and $\mathbf{a}'_j(u)=\mathbf{k}_j$ for $u> L+\frac{1+\dot{s}_j}{1-\dot{s}_1}$. Now, however, there is a segment between two kinks on each string and it propagates away from the junction as depicted in the right panel of Fig. \ref{examplefig}.

Calculation of the transmission coefficients \eqref{deftransmissioncoefficients} when the {\it first} kink encounters the junction yields
\begin{eqnarray}
C_1(\theta,\epsilon)&=&{\frac {(1-\epsilon)}{\sqrt{2\,\epsilon\,\cos  \theta
 +{\epsilon}^{2}+1}} 
} \; \; \leqslant 1 , \\
C_2(\theta,\epsilon)
&=& \frac{2\epsilon}{\sqrt{ (1+\epsilon+2\epsilon^2) + \epsilon(1-\epsilon)\cos\theta - (1+\epsilon)\sqrt{1-\epsilon^2}\sin\theta}} ,
\label{c2simp}
\\
C_3(\theta,\epsilon)&=&C_2(-\theta,\epsilon) \, .
\end{eqnarray}
In Fig. \ref{coeffsfig} we plot the transmission coefficients\footnote{We only plot $C_1$ and $C_2$, since $C_3 (\theta) = C_2 (-\theta)$ because of the symmetry of the initial configuration.} for different values of the string tensions in the allowed range $0 \leqslant  \epsilon=\mu_1/2\mu_2  \leqslant  1$.  One sees that, in this particular example, the reflected kink on string 1 always has a smaller amplitude than the incoming kink, though the reduction in amplitude is generally rather weak for incoming kinks on the lightest string.

By contrast, the transmitted kinks can be amplified. 
When all tensions are equal ($\epsilon = 1/2$), this occurs for a rather broad set of (static) junction configurations.
However for $\epsilon \rightarrow 0$ (that is, when strings 2 and 3 are heavy compared to string 1), we find $C_2>1$ only in a limited range of $\theta$. In this regime, it is in fact straightforward to understand the position of the peak. From (\ref{aprime}) it follows that the amplitude $A[{\ba'_j}]$ -- and hence the transmission coefficient $C_j$ -- is large when after the kink has crossed the junction, the corresponding $\dot{s}_j \rightarrow -1$.  Eq.~(\ref{sdot2}) predicts that, for small $\epsilon$,  this occurs on string 2 when $\sin \theta=1$ or $\theta=\pi/2$, as is indeed the case in Fig.~\ref{coeffsfig}. Away from its sharp peak, $C_2 \sim \epsilon$ for small $\epsilon$, as can be seen from eq.~(\ref{c2simp}).  Finally, for $\epsilon \sim 1$, $C_2$ is always close to 1 even though slight amplification can occur in a broad range of $\theta$.

\begin{figure}[H]
\begin{minipage}[t]{0.48\textwidth}
\centering
   \includegraphics[scale=0.35]{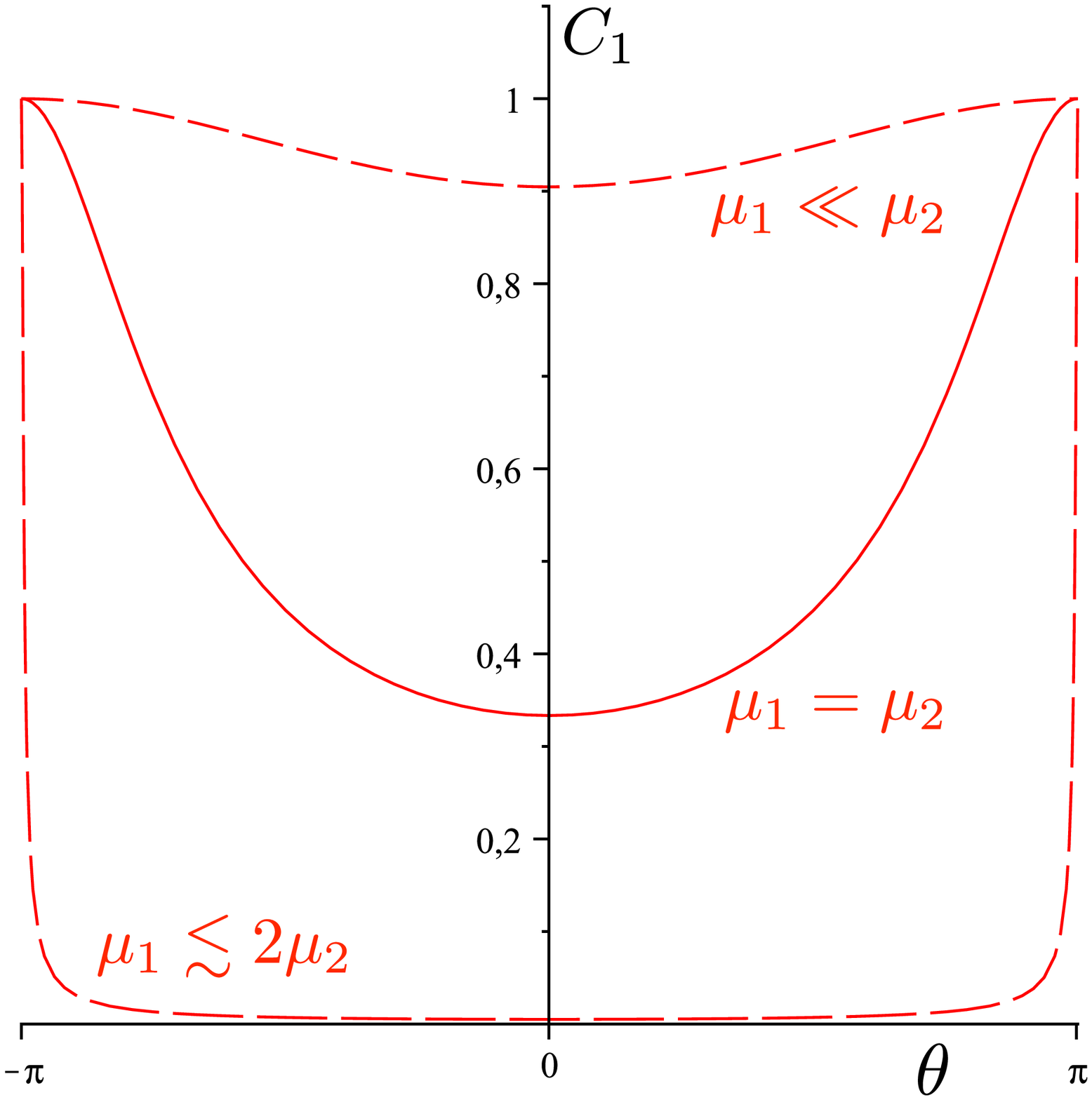} 
\end{minipage}
\begin{minipage}[t]{0.48\textwidth}
\centering 
   \includegraphics[scale=0.35]{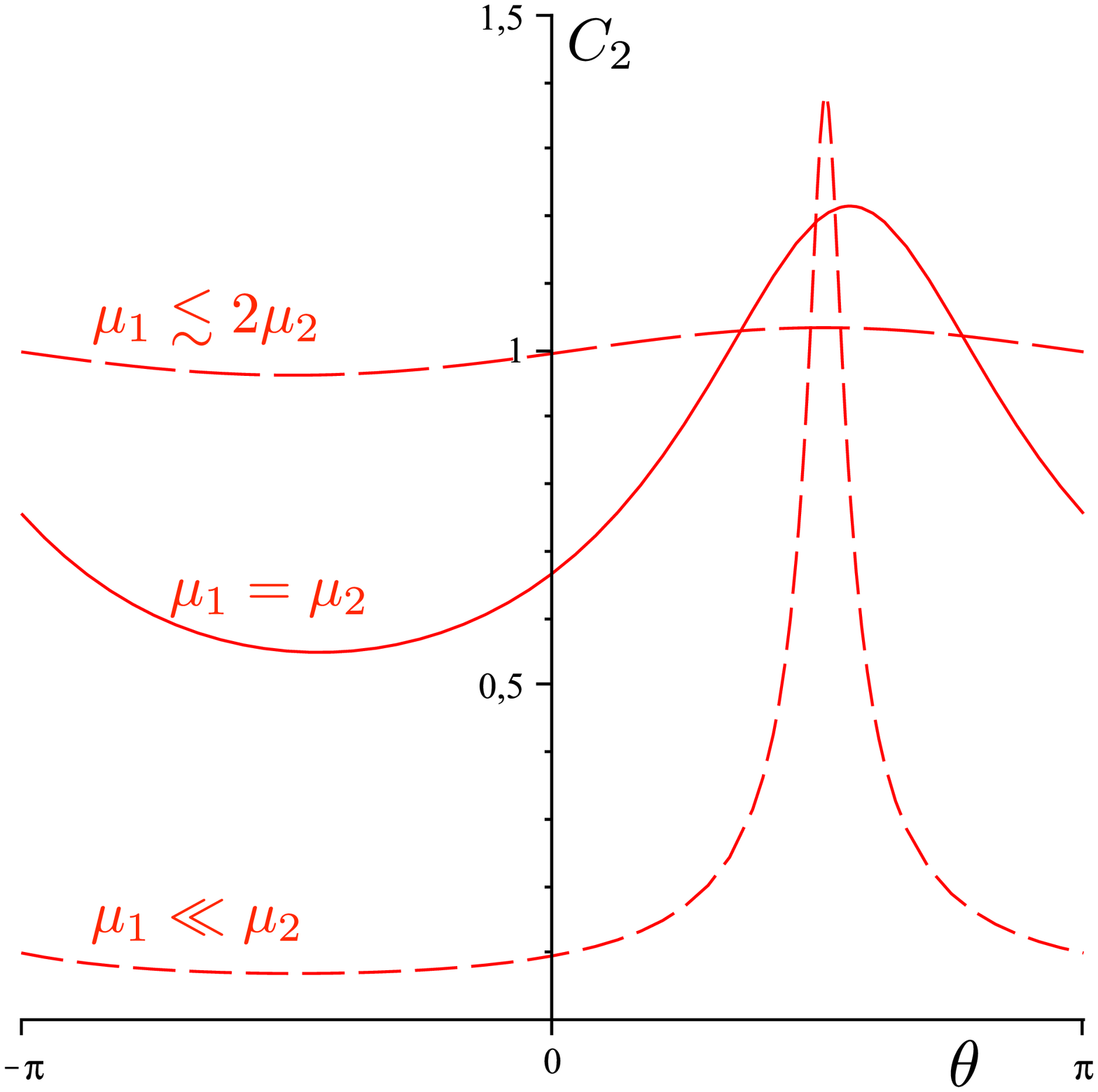} 
\end{minipage}
\caption{Transmission coefficients $C_1$ and $C_2$ as functions of $\theta$ for three different ratios of tensions $\mu_1$ and $\mu_2$. The tension $\mu_3$ of the third string is taken to be equal to $\mu_2$ in all three cases.}
\label{coeffsfig}
\end{figure}

We should stress that the above discussion is limited to the specific junction configuration of fig.~\ref{examplefig}. We now turn to arbitrary junction configurations.

\section{Distributions of transmission coefficients}
\label{sec:prob-dis}

In this section we aim to gain intuition on 
the distributions of transmission coefficients characterizing the propagation of kinks through junctions in a more generic context. Namely we study the statistical properties of the $C_j$ using an underlying uniform distribution of junction configurations.  We would like to answer the following question: is amplification frequent or not?

To specify the configuration of the junction just before the arrival of the incoming kink as well as the amplitude of the kink one needs 
four incoming unit vectors. We consider a kink moving towards the junction on string $1$, specified by a discontinuity in $\mathbf{b}'_1(v)$ at $v_{1*}$. Let $t_*$ be the time when the kink reaches the junction. Then the amplitudes of the transmitted kinks depend on $\mathbf{b}'_1(v_{1*}^+)$, $\mathbf{b}'_1(v_{1*}^-)$, $\mathbf{b}'_2(s_2(t_*)-t_*)$ and $\mathbf{b}'_3(s_3(t_*)-t_*)$, for which we will use the more concise notation $\mathbf{b}^{\prime +}_1$, $\mathbf{b}^{\prime -}_1$, $\mathbf{b}'_2$ and $\mathbf{b}'_3$. Eq.~(\ref{aprime}) then yields
\bea
\mathbf{a}^{\prime \pm} _1&=&P_1^\pm \mathbf{b}^{\prime \pm}_1
-Q_{1,2}^\pm \mathbf{b}'_2
-Q_{1,3}^\pm\mathbf{b}'_3\, ,
\label{a1}
\\
\mathbf{a}^{\prime \pm} _2&=&P_2^\pm \mathbf{b}'_2
-Q_{2,3}^\pm \mathbf{b}'_3
-Q_{2,1}^\pm\mathbf{b}^{\prime \pm}_1\, ,
\label{a2}
\\
\mathbf{a}^{\prime \pm} _3&=&P_3^\pm \mathbf{b}'_3
-Q_{3,1}^\pm \mathbf{b}^{\prime \pm}_1
-Q_{3,2}^\pm \mathbf{b}'_2\, ,
\label{a3}
\eea
where
\be
P_i^\pm = \left(\frac{1-\dot{s}^{\pm}_i}{1+\dot{s}^{\pm}_i}\right)\left(\frac{\nu_i}{\mu}\right),
\qquad
Q_{i,j}^\pm = \left(\frac{2\mu_j}{\mu}\right)\left(\frac{1-\dot{s}^{\pm}_j}{1+\dot{s}^{\pm}_i}\right)
\label{PQdef}
\ee
and the $\dot{s}_j^\pm$ are given by eq.~(\ref{spoint}).  Using eqs.~(\ref{a1}-\ref{a3}), the transmission coefficients $C_j$ can be obtained from (\ref{deftransmissioncoefficients}).  Our aim here is to determine their probability distributions.  Here we focus on the case of equal tensions. Unequal tensions are studied in Appendix A.

The configuration of the junction just before the arrival of the kink is specified by the unit vectors $\mathbf{b}^{\prime -}_1$, $\mathbf{b}'_2$ and $\mathbf{b}'_3$. We assume for now that these are independent, with uniform distributions on the unit sphere. We also assume a flat distribution in the incoming kink, namely a uniform distribution on the unit sphere for $\mathbf{b}^{\prime +}_1$ (and in particular independence of $\mathbf{b}^{\prime +}_1$ of the other unit vectors)\footnote{The consequences of the first hypothesis 
 have been studied in \cite{Copeland:2006if} where, for example, the probability distribution of the $\dot{s}^-_j$ was calculated. 
We will see in the next Sections that the second hypothesis does not hold for loops. It will turn out that most kinks on loops have a small amplitude. The flat distributions we use here should be interpreted as a working hypothesis adopted for now in the absence of a concrete dynamical model.}.

Given these assumptions, we numerically calculate the distributions of the transmission coefficient by drawing a large number (typically $N=3\times 10^7$) of random configurations at the junction in order to estimate the various statistical quantities of interest.

\subsection{Marginal Distributions $p(C_j)$}

We are in the first place interested in the joint distribution 
$p(C_1,C_2,C_3)$ from which one could determine, for example, the probability for several kinks to be amplified at the same time.
 However, as a warmup, we show in Figure \ref{alltensionsequal} the marginal distributions $p(C_j)$, for the case of equal tensions ${\mu_j=1}$.  These distributions $p(C_j)$ have a large tail where $C_j >1$, indicating that both the reflected and the transmitted kinks can be amplified in a significant part of configuration space.  Indeed, $P(C_1>1)=0.11$, $P(C_2>1)=0.19 = P(C_3>1)$.

 \begin{figure}[H] %  figure placement: here, top, bottom, or page
   \centering
   \includegraphics[scale=0.85]{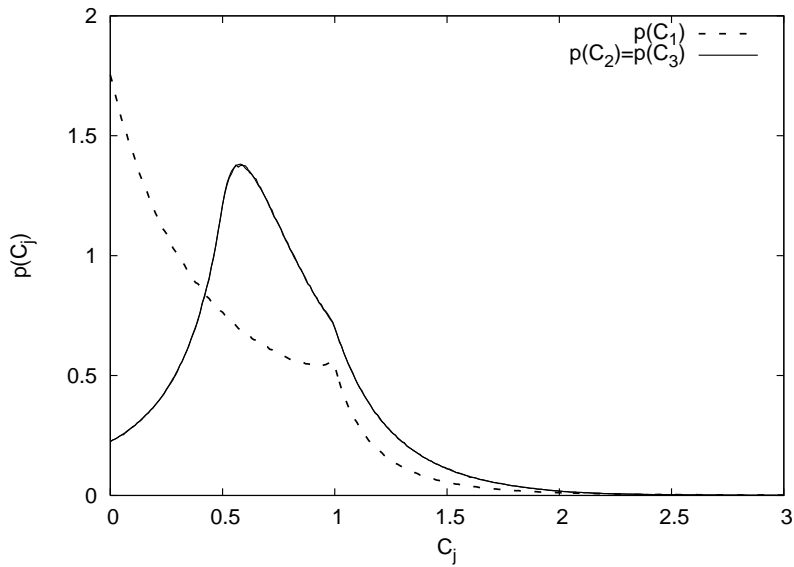} 
  \caption{Marginal distributions $p(C_j)$ of the transmission coefficients $C_j$ for equal tensions ${\mu_j=1}$ and a uniform distribution of junction configurations. 
    The average values are $\langle C_1\rangle =0.49$, $\langle C_2\rangle =0.72$, $\langle C_3\rangle =0.72$.} 
    %and $P(C_1<1)=0.89$, $P(C_2<1)=0.81$, $P(C_3<1)=0.81$}
   \label{alltensionsequal}
\end{figure}

However, as expected from the static example of Section \ref{example-static}, the marginal distributions depend significantly on the ratios of string tensions. In Appendix A we discuss the distributions for various other sets of tensions, including several limiting cases where, using the smallness of some of the coefficients (\ref{PQdef}), analytic arguments can explain certain features of the distributions.

\subsection{Simultaneous amplification of transmission coefficients}
\label{ssec:first}

In order to understand how the total number of large amplitude kinks on strings containing Y-junctions changes in time, one really needs the joint distribution $p(C_1,C_2,C_3)$ which contains information about the correlations between the different transmission coefficients.

The simplest question to ask involving correlations between the $C_j$ is the following: when a kink reaches a junction, what is the probability that {\it at least one} of the three daughter kinks is amplified (so at least one of the $C_j > 1$)? This probability may well be significantly larger than that suggested by the tails of the individual daughter kink distributions discussed above. 
In the case of equal tensions $\mu_i=1$ corresponding to Fig.~\ref{alltensionsequal} we find 
\beq
P(\text{at least one amplification})= 0.43 \qquad \text{for } \mu_j=1.
\eeq
Amplification is therefore not such a rare event, and hence we have a first hint that the number of \emph{large} amplitude kinks 
may grow significantly in a system in which the \emph{total} number of kinks increases due to the presence of Y-junctions.  

In a similar way, the probability of having {\it at least two} simultaneous amplifications is
\beq
P(\text{at least two amplifications})= 0.07 \qquad  \text{for } \mu_j=1.
\eeq
Again those probabilities are not negligible and such events can contribute importantly to the enhancement of the number of large amplitude kinks in an interconnected network\footnote{Numerically we have found no example of a simultaneous  {\it triple} amplification (no matter how large the number of random configurations generated). We have checked that $P(C_1\geq\alpha,C_2\geq\alpha,C_3\geq\alpha)$ is non-zero  any value of $\alpha$ strictly smaller than 1 (increasing the number of configurations always ends up yielding such an event) but this vanishies to zero when $\alpha=1$.}. In the table below, we summarize the amplification probabilities for the different sets of tensions considered here and in Appendix A.

{\footnotesize
\begin{center}
\begin{tabular}{|c|c|c|c|c|c|}
\hline
\hline
tensions & $P(C_1>1)$ & $P(C_2>1)$ & $P(C_3>1)$ & $P(\text{at least 1 amp})$ & $P(\text{at least 2 amp})$\\ \hline
$\mu_1=0.1$, $\mu_2=1$, $\mu_3=1$ & $0.12$ & $0.01$ & $0.01$ & $0.12$ & $0.004$\\ \hline
$\mu_1=1$, $\mu_2=0.1$, $\mu_3=1$ & $0.01$ & $0.19$ & $0.19$ & $0.65$ & $0.006$\\ \hline
$\mu_1=1.9$, $\mu_2=1$, $\mu_3=1$ & $0.12$ & $0.47$ & $0.47$ & $0.93$ & $0.11$\\ \hline
$\mu_1=1$, $\mu_2=1.9$, $\mu_3=1$ & $0.01$ & $0.49$ & $0.01$ & $0.19$ & $0.004$\\ \hline
$\mu_1=1$, $\mu_2=1$, $\mu_3=1$ & $0.12$ & $0.19$ & $0.19$ & $0.43$ & $0.07$\\ \hline
$\mu_1=1$, $\mu_2=1.2$, $\mu_3=1.4$ & $0.10$ & $0.11$ & $0.15$ & $0.31$ & $0.65$\\ \hline
\hline
\end{tabular}\\
\end{center}
}

%**********
%%

\subsection{Joint distribution $p(C_1,C_2,C_3)$}

Even though there is a significant region of configuration space in which the amplitude of at least one of the daughter kinks is enhanced relative to the amplitude of the initial kink, it is important to know whether the amplitude of the remaining kinks is typically significantly reduced in such events.
Fig.~\ref{slices} shows 2D slices through the joint distribution $p(C_1,C_2,C_3)$ for four different values of $C_1$, again with $\mu_j=1$. (Note that the grey scale differs from one panel to the next.)
The horizontal and vertical axes label $C_2$ and $C_3$ respectively.

\begin{figure}[h]
\begin{center}
$\begin{array}{c@{\hspace{-2cm}}c}
%\hspace{-2.5cm}
\includegraphics[scale=0.70]{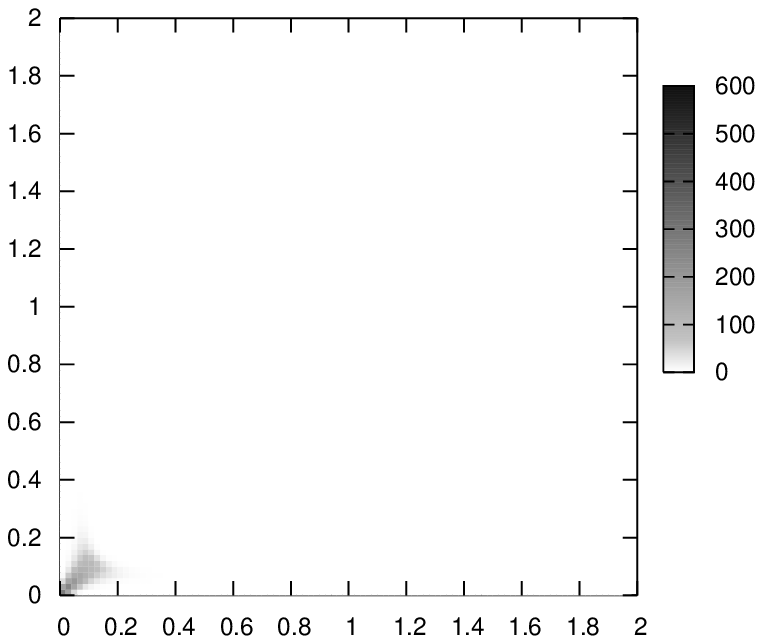} &
\includegraphics[scale=0.70]{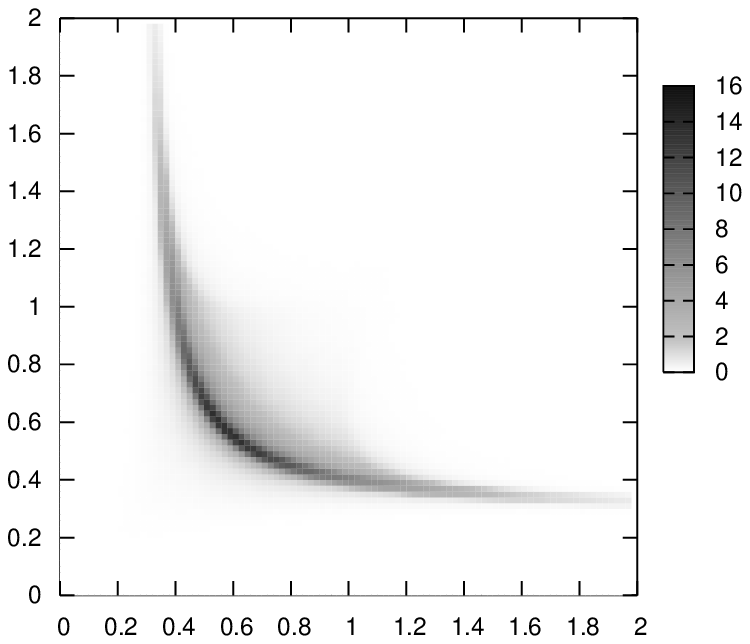} \\
%\hspace{-2.5cm}
\includegraphics[scale=0.70]{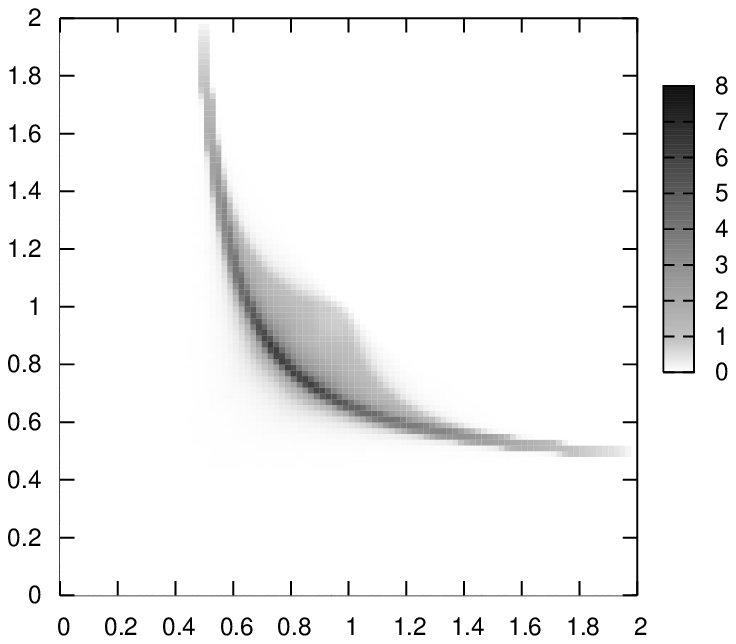} &
\includegraphics[scale=0.70]{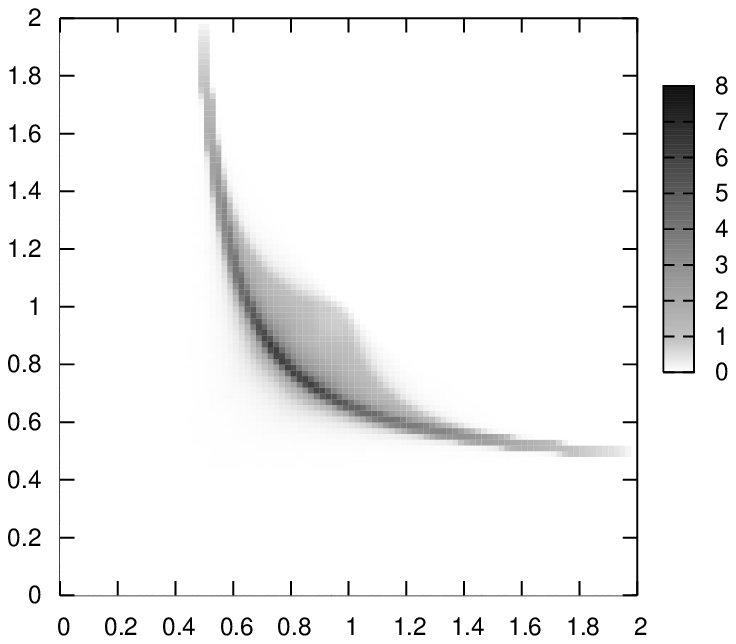} \\
\end{array}$
\end{center}
\caption{Slices of the joint distribution $p(C_1,C_2,C_3)$ for four different values of $C_1$ and $\mu_j=1$.  The horizontal and vertical axes label $C_2$ and $C_3$ respectively. The values of $C_1$ on the different slices are $C_1=0$ (top left), $C_1=0.2$ (top right), $C_1=1$ (bottom left) and $C_1=1.5$ (bottom right). Note that the grey scale differs from slice to slice.}
\label{slices}
\end{figure}

The joint probability distribution exhibits the following important features:
\begin{itemize}

\item The top left-hand panel shows that there is a sharp peak where all transmission coefficients take small values, $C_j \ll 1$.

\item 
The remaining three panels are for values of $C_1 \geq 0.2$.  They have a clear concentration of events on a arc-shaped line, as well as as events off that line at values of $C_2 \sim C_3$.  These latter events saturate around $C_2 \sim C_3 \lesssim1$ for $C_1 > 1$.

\item The arc-shaped lines finish on two ``amplification tails'' corresponding to values of $C_2$ or $C_3$ larger than $1$. 
There are no events for which $C_2\geq1$ and $C_3\geq1$ simultaneously.
\end{itemize}

Based on the above properties of the joint distribution one can conclude that the bulk of the transmission events falls in one of the following categories;
\begin{itemize}
\item All daughter kinks have drastically reduced amplitudes
\item All daughter kinks have slightly reduced amplitudes
\item The amplitude of the reflected kink is significantly reduced, one transmitted kink is amplified and one is slightly reduced
\item The reflected kink is amplified and the amplitude of the transmitted kinks is comparable to the amplitude of the incoming kink
\end{itemize}

We note in particular that the amplification of one of the daughter kinks does not imply that the amplitudes of the remaining daughter kinks are small. In fact, except for the first kind of transmission --- which is not considerably more frequent than the others, see Fig. \ref{alltensionsequal} ---  the amplitude of the scattered kinks is never strongly suppressed. This indicates that amplifications, combined with the rapid growth of the total number of kinks in a system with junctions, lead to a large number of ``large amplitude'' kinks. The next section is devoted to a quantitative study of this phenomenon.

\section{Proliferation of large amplitude kinks on a loop with junctions}
\label{section:idealised}

We have seen that the sharpness of the daughter kinks is generally comparable and occasionally larger than that of the incoming kink into a junction. Motivated by GW physics, it is therefore interesting to study how the number of kinks with a large amplitude (of order 1) evolves on closed loops with junctions. Indeed, a sustained growth of the number of such `large amplitude kinks' may well have an impact on the gravitational wave burst signal emanating from strings of this kind \cite{Binetruy10b}. 

The total number of kinks on a closed loop with junctions increases exponentially in time: when an initial kink reaches a junction it gives rise to three daughter kinks, which in turn propagate towards another junction where they multiply again, and so forth. Even though the amplitude of most of the daughter and higher generation kinks is small, it is clear that the number of large amplitude kinks will also grow exponentially {\it provided} amplification occurs sufficiently frequently. Here we show this is indeed the case in a simple model of a loop with Y-junctions, based as before on an underlying uniform distribution of junction configurations at the time of arrival of the kink. The evolution of the loop, for a relatively general class of initial loop configurations,  will be taken in account in section \ref{section:realistic} where we will see that the results obtained here remain largely valid.

\subsection{Setup}

The simplest example of a closed system with Y-junctions is a loop formed by three strings meeting at two junctions, a typical example of which is shown in Fig \ref{loop}. 

\begin{figure}[h] %  figure placement: here, top, bottom, or page
   \centering
   \includegraphics[scale=1]{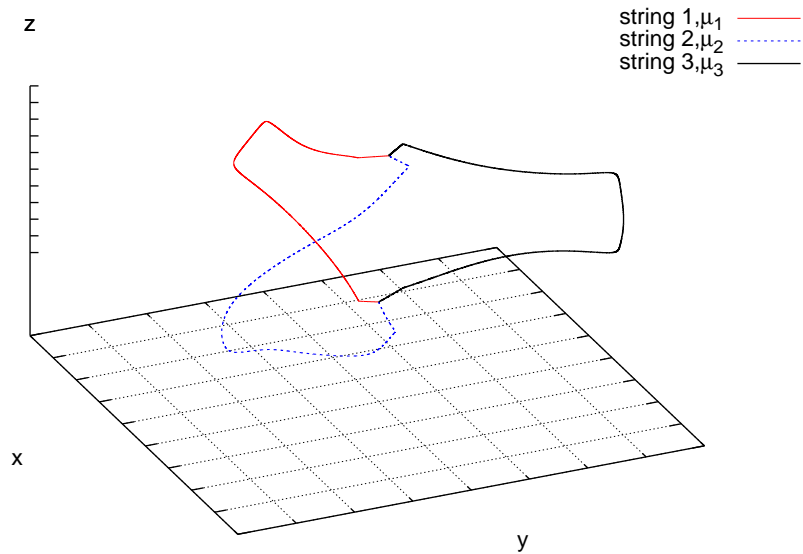} 
   \caption{{A typical loop formed by 3 strings and 2 junctions}}
   \label{loop}
\end{figure}

Our loop model is based on the following two simplifying assumptions;
\begin{enumerate}
\renewcommand{\labelitemi}{$\bullet$}
\item all three strings joining the two junctions have essentially the same \emph{constant} invariant length $L$;
\item when a kink (whose amplitude is known) reaches a junction, the configuration of the latter is randomly drawn among those that yield the correct amplitude for the original kink.
\end{enumerate}

More explicitly we proceed as follows.
First we specify the initial conditions at $t=0$, namely $K$, the total number of left moving and right moving kinks on all three strings, as well as their amplitudes. (The initial positions of those kinks do not enter the subsequent analysis.) 
After an interval of time $L$, the first assumption ensures that \emph{all} the initial kinks have reached a junction, but that \emph{none} of their descendants have done so (recall that kinks propagate at the speed of light on the strings). Therefore, at $t=L$, the system contains $3K$ kinks, and at time $t=nL$ with $n$ an integer, the total number of kinks in the system\footnote{In fact this is not exactly true if the constant invariant lengths of the strings are exactly equal. In this case, e.g. with $K=1$ the second generation kinks reach the second junction simultaneously and therefore recombine into a single kink per string instead of three. From then onwards, the total number of kinks in the system would remain constant and equal to three. We disregard this possibility and assume that, because of variations in the length of the strings when its evolutions is taken in account, the three kinks arrive at different times at the junctions giving rise to nine kinks.}  is $3^nK$. 
We then evaluate
\bea
Q^{A}_{j}(n)&=& \; \text {number of kinks on string } j \text{ of amplitude } \geq A 
\nonumber
\\
&& \text{as a function of the generation } n \text{, or equivalently time } t=nL. 
\nonumber
\eea

We do so as follows. Consider a given (say inward moving) kink of amplitude $A_i$ on string 1. Following assumption 2 we determine  the amplitude of the daughter kinks by randomly drawing junction configurations, similarly to Section \ref{sec:prob-dis}, but now
we restrict the choice to those configurations for which $\| \mathbf{b}^{\prime +}_{1}-\mathbf{b}^{\prime -}_{1}\|=A_i$. (In practice, we first draw $\mathbf{b}^{\prime -}_{1}$, $\mathbf{b}'_2$ and $\mathbf{b}'_3$ with uniform probabilities on the unit sphere and then draw $\mathbf{b}^{\prime +}_{1}$ with uniform probability on the intersection of the unit sphere and the sphere of center $\mathbf{b}^{\prime -}_{1}$ and radius $A_i$, \emph{i.e.} on a circle of radius $A_i$.) In other words, instead of using the probability distribution $p(C_1,C_2,C_3)$ to draw the transmission coefficients, we use an improved version $p_{A_i}(C_1,C_2,C_3)$ that takes into account the amplitude of the incoming kink.

To summarize, for each initial kink, our model amounts to building a tree of kinks. On this tree each node represents a kink, and contains the value of its amplitude as well as the number of the string on which it arrived. It has three daughter nodes whose values are drawn randomly according to the rules explained above. At the $n$th generation, the total number of kinks is $3^n$. Clearly, as there is no interaction between kinks, for $K$ initial kinks there are $K$ trees which evolve completely independently. Hence the statistical properties of systems originally containing several kinks can be trivially deduced from those by linearity. 

\subsubsection{Numerics}

Computationally, as the number of generations increases it becomes expensive to store $3^n$ amplitudes. To keep the computation manageable we divide the amplitude interval $[0,1]$ into $N_{bin}$ bins, and only keep track of the number of kinks on each string with an amplitude in the different bins, at each generation. To draw the amplitudes of the subsequent generation we use the center value of the bin as the amplitude of the incoming kink. That is, for kinks in bin $k >1$, we use as an amplitude $(k-\frac{1}{2})\frac{2}{N_{bin}}$. (In our simulations, we used $N_{bins}=100$). Finally, we set the amplitude of kinks in the first bin to zero: we expect that after a few generations most of the kinks that lie in this bin actually have an amplitude smaller than $1/N_{bin}$ by several orders of magnitude. Of course, this procedure can lead to an underestimation of the number of large amplitude kinks (because a few kinks from this bin must in reality be reamplified to yield large amplitude descendents) but it prevents many very small amplitude kinks from spuriously leading to large amplitude kinks. By doing this, we loose any information on the low amplitude part of the distribution (which we disregard in this paper). In our companion paper \cite{Binetruy10b}, we will refine this numerical setup to demonstrate that the large amplitude part of the distribution dominates the gravitational wave signal.

\subsection{Results}

In Fig \ref{expevol} we plot $\log Q^{1/4}_{1}(n)$  as a function of the generation $n$, for different sets of string tensions $\mu_1, \mu_2 \text{ and }\mu_3$.  (The particular choice of $A=1/4$ will be justified in \cite{Binetruy10b}.) The initial condition at the start of the simulation was a single right-moving kink $K=1$ of maximal amplitude on each string.

\begin{figure}[H]
\centering
\includegraphics[scale=1]{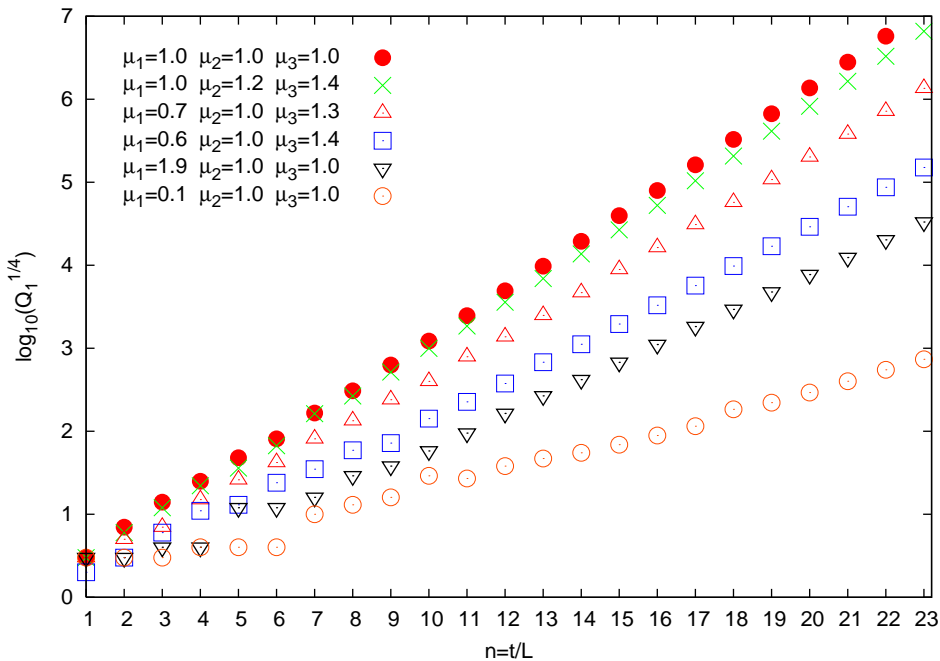} 
\caption{The logarithm of the number of kinks on string $1$ of amplitude $A \geqslant 1/4$, as a function of time measured in units of the constant string length $L$, for different sets of string tensions $\mu_1, \mu_2 \text{ and }\mu_3$. }
  \label{expevol}
  \end{figure}
After some initial fluctuations at small $n$, which vary significantly from one realization to another\footnote{This is because as long as the number of kinks is small, the random values of the $C_j$ that are drawn crucially affect the distribution of kinks.}, one sees that the points corresponding to a given set of string tensions can be fitted to a straight line, whose slope is independent of the particular realization. Hence $Q^{1/4}_{1}(n)$ grows exponentially with $n$. More generally we find\footnote{Empirically we find the slope of $\log Q^{A}_{j}(n)$ is independent of $j$ for large $n$.}
\begin{equation} \label{defQ}
Q^{A}_{j}(n)\propto \exp[\gamma n]
\end{equation}
where the coefficient $\gamma$ depends on the tensions $\mu_j$ as well as on the amplitude $A$. One sees proliferation is most efficient when all tensions are equal.  Gradually moving away from this case, the slope $\tilde{\gamma}=\gamma/\ln(10)$ of the curves in Fig \ref{expevol} decreases and approaches zero 
when one or more of the $\nu_j$ given in (\ref{nudef}) vanish. For a given ratio of tensions, we find $\gamma$ is approximately independent of $A$ over most of the range of possible amplitudes. For equal tensions and $A \geq 0.1$ shown in Fig \ref{expevol} one has $\tilde{\gamma} \approx 1/3$. In the limit $A \rightarrow 0$, the slope sharply increases to $\gamma \approx \ln 3$, since obviously $Q^{0}_{j}(n)=3^n$.

We therefore conclude that, at least in this simplified model of a loop with junctions, for a large range of string tensions, the amplification rate is sufficient to sustain an exponential growth of the number of large amplitude kinks. In appendix B we illustrate, with a toy analytical model, the origin of this exponential growth.
We must emphasize, however, that since amplification remains a rare event, the vast majority of kinks at sufficiently late times will have small amplitude. Indeed, the fraction of the total number of kinks that have a large amplitude tends to zero.  The implications of these findings for the gravitational wave signal emitted by string loops of this kind will be studied elsewhere \cite{Binetruy10b}.

\section{Proliferation of large amplitude kinks on evolving loops with junctions}
\label{section:realistic}

The model of a loop with junctions discussed in section \ref{section:idealised} does not take in account the  dynamics of the loop. In particular, it assumes the invariant length of the strings forming the loop is constant in time.
Here we include the effects of the loop dynamics on the evolution of the number of large amplitude kinks, by numerically integrating the equations of motion \eqref{spoint} and \eqref{aprime}, suitably modified to take account of the presence of the two junctions.

This is more complicated than for periodic loops with no junctions, since one needs to keep track of the position of each junction, and extend the definition of the $\ba_j'$ beyond their initial domain. 
Our simulations generalize those of \cite{Bevis:2009az} in which certain initially static and symmetric planar loop configurations were studied. Here we consider a rather general class of non-static initial conditions, and count the number as well as the amplitude of the kinks on the loop as a function of time. Note, however, that we do {\it not} take in account self-intersections between the strings in our simulations. This may be an important limitation of our model if the probability of intercommutation is large.  (In this case, intercommutation between strings of the same type would lead to a loop being chopped off, whereas intercommutation between strings of different types might increase the number of junctions in the loop.)  Our simulations end when the length of one of the strings connecting the two junctions shrinks to zero and the junctions collide. The outcome of such a collision is an open question, but it may well lead to the formation of two loops without junctions (assumed for e.g.~in \cite{Hassan1}): this would end the proliferation of kinks.

\subsection{Initial conditions}

Our initial condition consists of 3 segments of string of initial invariant length $L_j(t=0)$ (specified below) which join two junctions at positions
\be
\bX_A = (0,0,-1) \qquad \bX_B=(0,0,1).
\ee
Motivated by the harmonic construction of standard cosmic string loops with no junctions \cite{KT82,Tur84,CDH,DES,Siemens:1994ir}, the three initial string segments are taken to be three arcs of circles to which we add a higher harmonic. 
%Each string is initially contained in a vertical plane obtained by rotation of (xOz) about the z axis. (***********)  
Explicitly, 
\begin{equation}
\bx_j(\sigma,t=0)=
\mathcal{R}(\alpha_j)
\left(
\begin{array}{c}
-\sin(\frac{\sigma}{H_j})\left[1+a_j\sin\left(m_j\frac{\sigma}{H_j}\right)\right]\\
0\\
\cos(\frac{\sigma}{H_j})\left[1+b_j\sin\left(n_j\frac{\sigma}{H_j}\right)\right]\\
\end{array} \right)
\end{equation}
where $m_j,n_j \in \mathbf{N}$, and $\mathcal{R}(\alpha_j)$ is the rotation matrix about the $z$ axis by an angle $\alpha_j$
\begin{equation}
\mathcal{R}(\alpha_j)=
\left(
\begin{array}{ccc}
\cos(\alpha_j) & -\sin(\alpha_j) & 0\\
\sin(\alpha_j) & \cos(\alpha_j) & 0\\
0 & 0 & 1
\end{array} \right).
\label{initial}
\end{equation}
If we set $a_j=b_j=0$, then the segments are initially semi-circles. If $a_j$ or $b_j$ are nonzero then the strings appear as perturbed semi-circles (at least for small values of $a_j$ and $b_j$), and the integers $m_j$ and $n_j$ set the `wiggliness' of the perturbation.
The normalisation 
\begin{equation}
H_j=\sqrt{(1+a_j)^2+(a_jm_j)^2+(1+b_j)^2+(b_jn_j)^2}
\end{equation}
ensures that $\|\mathbf{x}'_j\| \leq1$ since $\dot{\bx}_j^2 + \bx_j^{\prime 2} = 1$ from the gauge conditions \eqref{conf-gauge}.  Note that with this definition of $H_j$ in general $\sup \|\mathbf{x}'_j \| <1$ so that the initial condition is never static. The initial invariant length of the strings segments is $L_j(0)=\pi H_j$ so that
\begin{equation}
\sigma\in[-\pi H_j, 0] \; \; {\rm at} \; \; t=0.
\end{equation}
In order to compare our results with Section 4, we will measure time in units of 
\be
L=\frac{1}{3}(L_1(t)+L_2(t)+L_3(t)),
\label{Ldef}
\ee
 the total invariant length of the loop divided by the number of string segments. Note that this quantity remains constant throughout the evolution.

The initial velocity of the strings $\dot{\bx}_j$ must satisfy the gauge condition $\dot{\bx}_j \cdot \bx'_j=0$. We choose it to be
\begin{equation}
\dot{\mathbf{x}}_j(\sigma,t=0)=
N_j(\sigma)\mathcal{R}(\alpha_j)
\left(
\begin{array}{c}
\sin(\frac{\sigma}{H_j})\left(1+b_j\sin\left(n_j\frac{\sigma}{H_j}\right)\right)-b_jn_j\cos(\frac{\sigma}{H_j})\cos(n_j\frac{\sigma}{H_j})\\
v_j\\
-\cos(\frac{\sigma}{H_j})\left(1+a_j\sin\left(m_j\frac{\sigma}{H_j}\right)\right)-a_jm_j\sin(\frac{\sigma}{H_j})\cos(m_j\frac{\sigma}{H_j})\\
\end{array} \right)
\end{equation}
with $N_j(\sigma)$ defined so that $\dot{\textbf{x} } _{j} ^{2}=1-\textbf{x}  _{j}  ^{\prime 2}$, and $v_j$ is the component of velocity transverse to the plane of the string.
Finally the functions $\mathbf{a}'_j(z)$ and $\mathbf{b}'_j(z)$ on the interval $[-\pi H_j,0]$ are obtained through
\begin{eqnarray}
\mathbf{a}'_j(z)&=&\mathbf{x}'_j(\sigma=z,t=0)+\dot{\mathbf{x}}_j(\sigma=z,t=0)\\
\mathbf{b}'_j(z)&=&\mathbf{x}'_j(\sigma=z,t=0)-\dot{\mathbf{x}}_j(\sigma=z,t=0).
\end{eqnarray}

As explained in Section \ref{sec:kinkamp}, the $\sigma$ parameter on each string takes values in $[s_{A,j}(t),s_{B,j}(t)]$ at time $t$. Furthermore, integrating the equations of motion extends the definition of the functions $\mathbf{a}'_j(z)$ and $\mathbf{b}'_j(z)$ outside the interval $[-\pi H_j,0]$ (using equations \eqref{aprime}) to $z>0$ for $\mathbf{a}'_j(z)$ and to $z<-\pi H_j$ for $\mathbf{b}'_j(z)$. Thus if the evolution is calculated up to a final time $t_f$, then at the end of the simulation $\mathbf{a}'_j(z)$ will be defined in the interval $[-\pi H_j,z_{f,j}]$ where $z_{f,j}=t_f+s_{B,j}(t_f)$.

\subsection{Proliferation of large amplitude kinks}

This class of initial conditions has $6$ parameters for each string: $\alpha_j$, $a_j$, $b_j$, $m_j$, $n_j$ and $v_j$, and therefore enables one to probe a variety of initial configurations. 
We now evolve these strings and count the number of large amplitude kinks as a function of time. That is, we calculate $A[\ba'_j](z)=\frac{1}{2} \|\mathbf{a}'_j(z^+)-\mathbf{a}'_j(z^-)\|$ (for left-moving kinks) and $A[\bb'_j](z)=\frac{1}{2} \|\mathbf{b}'_j(z^+)-\mathbf{b}'_j(z^-)\|$
as functions of $z$. These functions are zero except at the position of a kink, where they reduce to the kink amplitude.

As discussed in section \ref{sec:kinkamp}, even though the initial configuration is infinitely smooth and appears to contain no kinks, this is not the case.  While $\mathbf{a}'_j$ and $\mathbf{b}'_j$ are continuous \emph{inside} the interval $]-\pi H_j,0[$ they have a discontinuity at $z=-\pi H_j$ for $\mathbf{b}'_j(z)$ and $z=0$ for $\mathbf{a}'_j(z)$. Indeed, as soon as the evolution starts, the equations of motion define the function $\mathbf{a}'_j(z)$ for $z>0$. In particular, $\mathbf{a}'_j(z=0^+)$ depends only on the values of the $\mathbf{b}'_\ell(z=0^-)$ for $\ell=1,2,3$ and differ from $\mathbf{a}'_j(z=0^-)$. 
Our loop therefore initially contains $6$ kinks: one left-moving and one right-moving on each string of the loop.

The numerical integration of the equations of motion ends
at time $t_f$ when the two junctions collide.
The results are shown in figure \ref{evolnumberkinksreal} for the case of equal tension strings $\mu_j=1$. Here we have used the following set of parameters as initial conditions: string 1 $(\alpha_1=0,A_1=.2,B_1=.3,m_1=2,n_1=3,v_t=0)$, string 2 $(\alpha_2=2\pi/3,A_2=.1,B_2=.2,m_2=3,n_2=4,v_t=0)$ and string 3 $(\alpha_3=4\pi/3,A_3=.2,B_3=.4,m_3=1,n_3=3,v_t=0)$. A snapshot of this loop shortly after the beginning of the simulation is shown in Figure \ref{loop}. One can see the six kinks propagating away from the junctions.

\begin{figure}[H]
\centering
\includegraphics[scale=1]{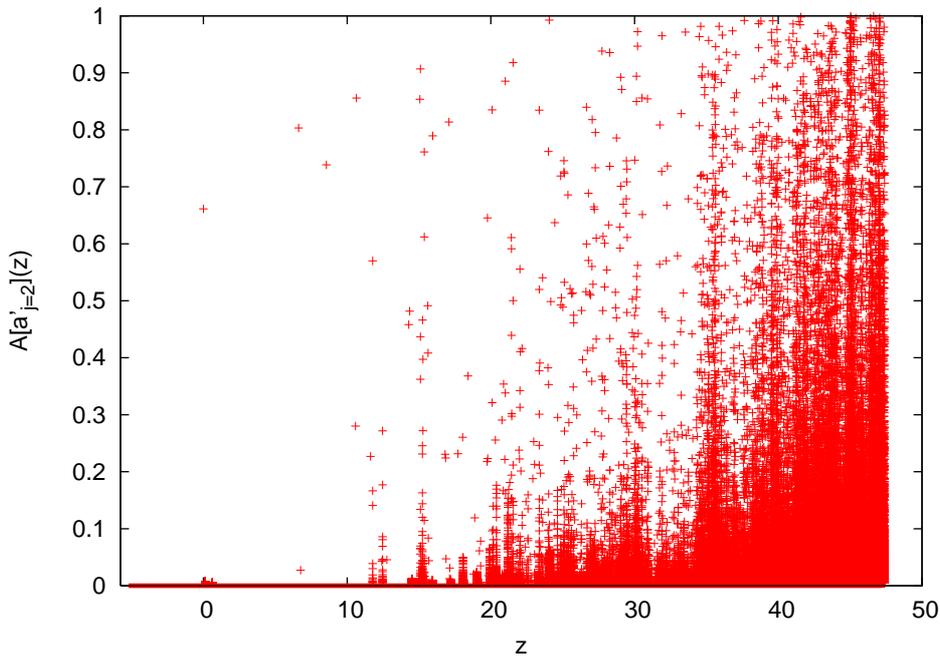} 
\caption{$A[\ba'_{j=2}](z)$ as a function of $z$. We evolved until $t_f=50$ which corresponds to $z_{f,2}=47.4$.
Each point above zero corresponds to a left moving kink propagating on string number $2$. As the system evolves, many kinks are created and although the vast majority has a very small amplitude, a large number has an amplitude of the order of $1$.}
\label{evolnumberkinksreal}
\end{figure}

As one can see in Fig.~\ref{evolnumberkinksreal} the total number of (left moving kinks) that have propagated on string number $2$ is very large, and even though the amplitude of many of those kinks is small there still is a large number of those kinks with an amplitude larger than e.g.~$1/4$. This is in line with the results in the previous section. We can be more precise by calculating $Q_j^{1/4}$, the number of (left moving) kinks propagating on string $j$ with an amplitude larger than $1/4$, as a function of time measured in units $L$ defined in (\ref{Ldef}). The result is shown in Fig.~\ref{kinkrealstring2}. One clearly sees that 
after some initial fluctuations, a regime of exponential growth sets in.

Remarkably, the slope of the curves in Fig.~\ref{kinkrealstring2} is larger than the slope of the corresponding function $\log Q_j^{1/4}$, shown in Fig.~\ref{expevol} for the simple model of a loop with junctions discussed in Section 4 (where we do do not take in account the loop evolution). 
The reason that loop dynamics has this effect on the proliferation process is simply because the loop  dynamics generally implies that one of the strings shrinks. On this string, kinks propagate more frequently back and forward between the junctions, thereby increasing the rate of proliferation. One expects therefore that, for a given set of tensions, proliferation is in fact the least efficient when all lengths are constant and equal, which is just the case considered in Section 4, and that the rate obtained here is more realistic. We note when the string that is shrinking becomes small, kinks tend to have a smaller amplitude on it.

\begin{figure}[H]
\centering
\includegraphics[scale=1.1]{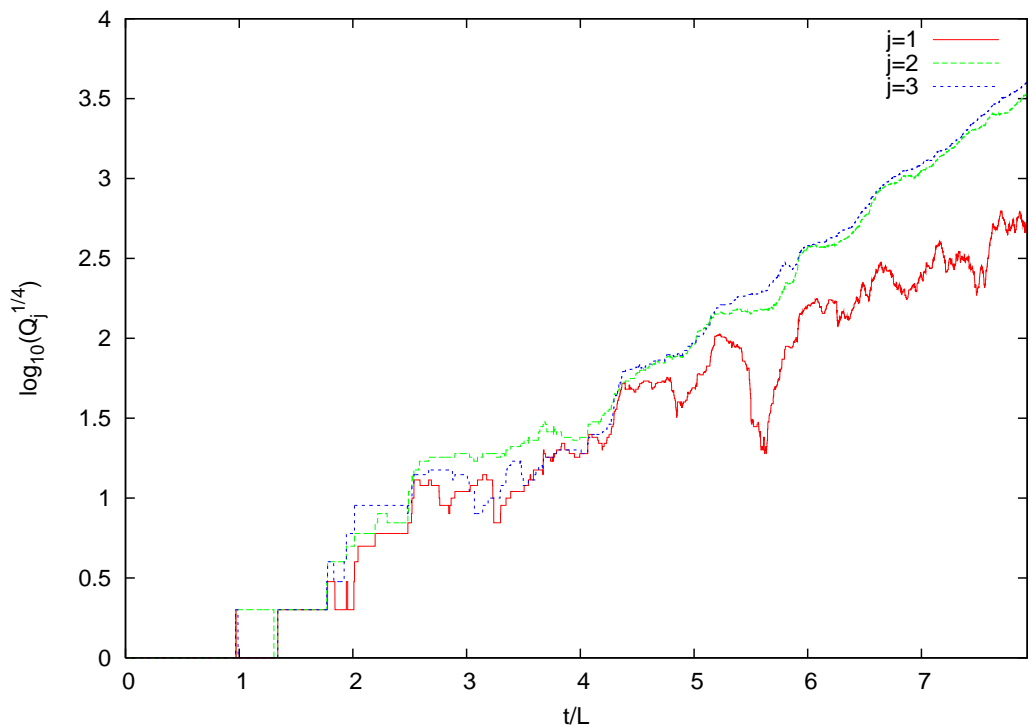} 
\caption{The logarithm of the number of (left moving) kinks of amplitude $A \geq 1/4$ on each string, as a function of time measured in units of the total invariant length of the loop $L$ defined in \eqref{Ldef} (which remains constant in time), for the case of strings with equal tensions. After some initial fluctuations, a regime of exponential growth sets in.}
\label{kinkrealstring2}
\end{figure}

Finally, we note that these simulations end at $t_f=50$ because the junctions collide briefly after that time, as is evident from figure \ref{invlengths}. At that time, the total number of kinks on this loop is of the order $10^4$.

\begin{figure}[H]
\centering
\includegraphics[scale=1]{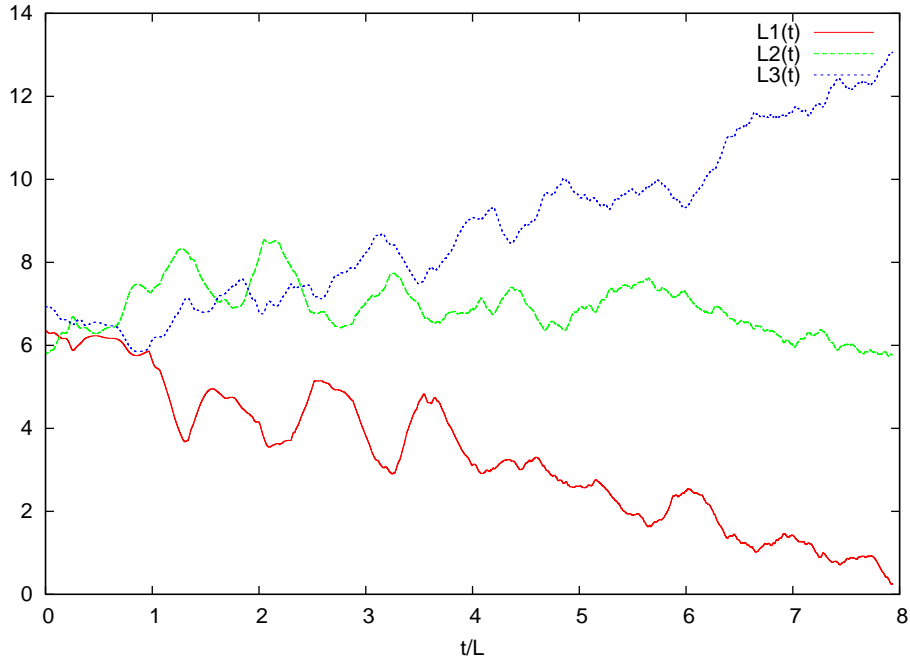} 
\caption{Evolution of the "Invariant lengths" of the strings. The simulations end around $t_f =50$ when string 1 shrinks to a point and the junctions collide.}
\label{invlengths}
\end{figure}

\subsection{Discussion}

The details of the evolution of the number of kinks evidently depend somewhat on the initial conditions. However, our central result that the number of large amplitude kinks proliferates exponentially appears to be universal. In particular, it is a robust feature of the evolution for the wide range of parameter values we have scanned.

The loop evolution has two important implications for the evolution of kinks. First, as discussed above, it enhances the proliferation rate because the length of one of the strings generally decreases. On this string, kinks propagate more frequently back and forward between the junctions.

However, at the same time this possibly provides an end to the proliferation process. 
Indeed, for all the initial conditions that we have tried, one of the strings always 
ended up shrinking to a point, resulting in the collision of the two junctions. The result of such a collision is unclear and depends on the physics of the underlying theory \cite{Bevis:2009az}. The junctions might disappear thus ending the proliferation, or two new junctions might form and proliferation may then continue.

Gravitational backreaction (e.g. by rounding off the kinks or by inducing the decay of the loop due to important radiation from the many kinks) may also end the proliferation. Finally, radiation of other fields might also become important as the number of kinks increases and could also play a limiting r\^ole.

\section{Conclusion}
\label{conc}

Motivated by their effects on the gravitational wave emission of cosmic strings and superstrings, we have studied the dynamics of kinks on strings with junctions. We have concentrated in particular on the evolution of the number of very sharp -- or equivalently, large amplitude -- kinks since it turns out these provide the dominant contribution to the GW burst signal from kinks on a network of strings  \cite{Binetruy10b}.

The propagation of a kink through a Y-junction leads to the formation of three daughter kinks -- one reflected kink and two transmitted kinks. We first showed analytically that, for a specific initially static junction configuration, one or two of the daughter kinks can be sharper than the incoming kink. This turns out not to be an isolated case: the amplification of kinks through their interaction with junctions is a rather generic phenomenon. Indeed we showed in Section \ref{sec:prob-dis} that, assuming a random distribution for the four incoming waves specifying the junction configuration and for equal string tensions, kinks are amplified in a significant region of configuration space. In appendix A we have generalized this calculation to strings of different tensions finding similar results.

The dominant contribution to the GW signal from a network of strings comes from the loops. In Sections \ref{section:idealised} and \ref{section:realistic} we have therefore studied the evolution of kinks on loops with junctions. We have considered loops which, for simplicity, contain 2 junctions. If one neglects the loop dynamics and assumes that {\it i)} all strings joining the two junctions have essentially the same invariant length, and also that {\it ii)} when a kink of known amplitude reaches a junction, the configuration of the latter is randomly drawn among those that yield the correct amplitude for the original kink, then one finds the amplification rate is sufficient to sustain an exponential growth of the number of large amplitude kinks. For a wide range of tensions, the coefficient in the exponent appears to be of order one when time is measured in units of $L$. The origin of this exponential growth was illustrated with a toy model in Appendix B.

We have included the effect of the loop dynamics on the proliferation and amplification of kinks in Section \ref{section:realistic}, where we numerically integrated the equations of motion of a loop with two junctions for a rather general class of initial conditions. Our simulations generalize those of \cite{Bevis:2009az} who considered an initially static and symmetric planar loop configuration. Interestingly, we find the loop evolution somewhat enhances the rate at which the number of large amplitude kink grows. To a large extent this difference can be traced to the fact that under evolution, generically one of the strings shrinks. On this string, kinks propagate more frequently back and forward between the junctions, thereby increasing the rate of proliferation and thus also the number of large amplitude kinks. By the time the junctions on the loop collide we typically find at least as many as $\sim 10^4$ large amplitude kinks in the equal tension case.

We note, however, that our simulations do not take into account intercommutations between strings. These may lead to the creation of new junctions as well as several smaller loops (with many kinks). Further, we end our simulations just before the junctions collide and have not addressed the subsequent evolution of the system. Finally, we note that backreaction effects will become increasingly important as the number of kinks grows, and that our simulations do not include this effect.

Nevertheless our findings suggest that if an evolving network of strings with junctions contains a population of loops with junctions, these typically contain a large number of very sharp kinks. The implications of this for the GW signal emitted by networks of this kind will be discussed elsewhere \cite{Binetruy10b}.

\section*{Acknowledgements}

DAS thanks Tom Kibble, Hassan Firouzjahi , Christophe Ringeval and Tanmay Vachaspati for useful comments and discussions. This work was supported by the Agence Nationale de la Recherche grant "STR-COSMO" (ANR-09-BLAN-0157).

\section*{Appendix A}
\label{appA}

In this Appendix we study the dependence of the distributions $p(C_j)$ of transmission coefficients on the 
ratios of the tensions of each of the strings, for three semi-infinite strings meeting at a junction.
The case of equal tensions was discussed in the text in Section 3 where it was found that the distributions of both the reflected and the transmitted kinks have significant tails where $C_j >1$. As in Sec 3, we consider a kink moving towards the junction on string $1$ and assume a uniform distribution on the unit sphere of the vectors $\mathbf{b}'_j$ that specify the junction configuration.

\subsection*{Incoming kink on a light string: $\mu_1 \ll \mu_2 \sim \mu_3$}

The kink reaches the junction on a string which is much lighter than the other two strings.  One might therefore expect, on average, the kinks transmitted to the heavier strings to have reduced amplitudes and the reflected kink to have an amplitude comparable or even enhanced relative to the amplitude of the incoming kink.  

This is indeed what we find. Figure \ref{mu1light} shows the marginal distributions $p(C_j)$ for strings of tensions ${\mu_1}=0.1$, ${\mu_2=\mu_3=1}$.  
One sees the distribution of $C_1$ is relatively flat, with a probability of amplification $P(C_1>1)$ larger than ten per cent. On the other hand $p(C_2)$ is sharply peaked around a value much smaller than one.

The existence of this peak can be understood from eq.~(\ref{a2}). Let $\epsilon={\mu_1}/{\mu}\ll1$ and take 
$\mu_2\sim \mu_3$ so that $\nu_1/\mu \sim O(\epsilon)$. Then from \eqref{energy}
it follows that $\dot{s}_2^\pm+ \dot{s}_3^\pm= O(\epsilon)$ so that
\begin{equation}
\mathbf{a}^{\prime \pm}_2=\left(  \frac{1-\dot{s}^\pm_2}{1+\dot{s}^\pm_2}\right) \underbrace{\left( \frac{\nu_1}{\mu}\right)}_{O(\epsilon)}\mathbf{b}'_2
-\underbrace{\left( \frac{2\mu_3}{\mu}\right)}_{1+O(\epsilon)}\underbrace{\frac{1-\dot{s}^\pm_3}{1+\dot{s}^\pm_2}}_{1+O(\epsilon)} \mathbf{b}'_3
-\underbrace{\frac{2\mu_1}{\mu}}_{O(\epsilon)}\frac{1-\dot{s}^\pm_1}{1+\dot{s}^\pm_2} \mathbf{b}^{\prime \pm}_1.
\end{equation}
Now generically one has $1 - \dot{s}_2^\pm \sim O(\epsilon)$ for the above configuration of tensions
\cite{Copeland:2006if}.
Therefore, since the functions $\bb'_{2}$ and $\bb'_{3}$ do not change when the kink crosses the junctions, it follows that
\be
A[\ba'_2] = \frac{1}{2} \| \ba^{\prime -}_2 - \ba^{\prime +}_2 \| \sim O(\epsilon) A[\mathbf{b}'_1] \qquad \Longrightarrow \qquad C_2 \sim O(\epsilon) .
\ee
The coefficient $C_2$ will be large only if either $\dot{s}_2^{+}$ or  $\dot{s}_2^{-}$ is close to minus one.
However the probability for this is small; according to \cite{Copeland:2006if} only three per cent of the $\dot{s}_2^\pm$ lie in the interval $[-1,-0.9]$, which explains why very few events are seen in the tail at large values of the distributions of $C_2$ and $C_3$.

Finally, the average values of the transmission coefficients for the above set of tensions are given by
\be
\langle  C_1\rangle =0.49, \qquad \langle  C_2\rangle =\langle  C_3\rangle = 0.09.
\ee

\hspace{-.5cm}
\begin{minipage}[t]{0.48\textwidth}
 \begin{figure}[H]
   \centering
   \includegraphics[scale=0.65]{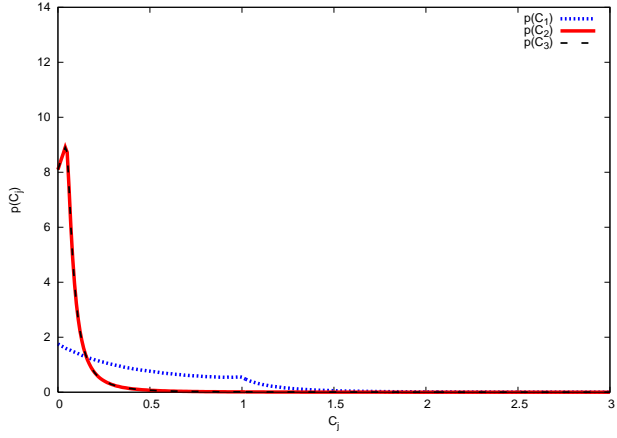} 
   \caption{Distributions of the different transmission coefficients for tensions $\mathrm{\mu_1=0.1}$, $\mathrm{\mu_2=\mu_3=1}$, with the incoming kink on string 1. The average values are $\langle  C_1\rangle =0.49$, $\langle C_2\rangle =\langle C_3\rangle =0.09$ and $P(C_1<1)=0.88$, $P(C_2<1)=P(C_3<1)=0.99 $.}
  \label{mu1light}
  \end{figure}
\end{minipage}
\hspace{.5cm}
\begin{minipage}[t]{0.48\textwidth}
 \begin{figure}[H]
   \centering
   \includegraphics[scale=0.65]{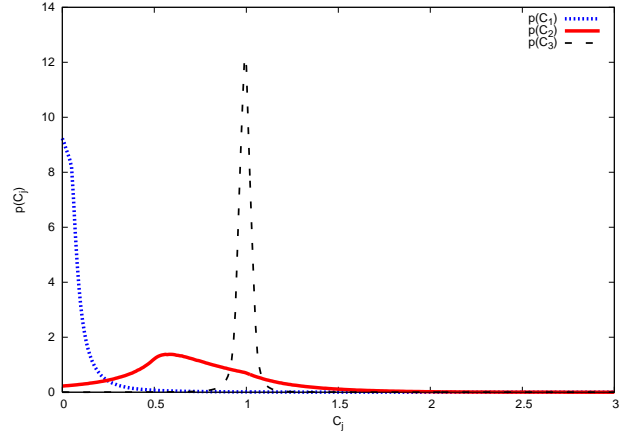} 
   \caption{The same distributions as in Fig \ref{mu1light} but for tensions $\mathrm{\mu_1=1}$, $\mathrm{\mu_2=0.1}$ and $\mathrm{\mu_3=1}$. Here $\langle C_1\rangle =0.09$, $\langle C_2\rangle =0.72$, $\langle C_3\rangle =0.99$ and $P(C_1<1)=0.99$, $P(C_2<1)=0.81$, $P(C_3<1)=0.53$.}
  \label{mu2light}
  \end{figure}
\end{minipage}\\

\subsection*{Incoming kink on a heavy string: $\mu_1 \sim \mu_3 \gg \mu_2 $} 

In this case, the kink propagates on a heavy string towards a junction consisting of another heavy string as well as a very light one. The light string is expected to play a minor role while the heavy strings almost behave as a single long string without junction on which the kink simply propagates. Hence the reflected kink amplitude $C_1$ should be peaked about a very small value, while the transmitted kink amplitude on the heavy string $C_3$ should be peaked about 1.  This is indeed what we find, as shown in Fig \ref{mu2light}), and an analytic argument similar to the one given above can be used to explain the position of the peaks in the distributions\footnote{The amplitude of the kink transmitted to the light string depends crucially on the configuration, as indicated by the relatively flat distribution of $C_2$.} of $C_1$ and $C_2$.

We now have $\dot{s}_1+ \dot{s}_3= O(\epsilon)$. The argument for $C_1$ is exactly the same as above,
since
\begin{equation}
\mathbf{a}'_1=\frac{1-\dot{s}_1}{1+\dot{s}_1}\underbrace{\left(1-\frac{2\mu_1}{\sum_k \mu_k}\right)}_{O(\epsilon)}\mathbf{b}'_1
-\underbrace{\frac{2\mu_2}{\sum_k \mu_k}}_{O(\epsilon)}\frac{1-\dot{s}_2}{1+\dot{s}_1} \mathbf{b}'_2
-\underbrace{\frac{2\mu_3}{\sum_k \mu_k}}_{1+O(\epsilon)}\underbrace{\frac{1-\dot{s}_3}{1+\dot{s}_1}}_{1+O(\epsilon)} \mathbf{b}'_3
\end{equation}
For $C_3$, we have
\begin{equation}
\mathbf{a}'_3=\frac{1-\dot{s}_3}{1+\dot{s}_3}\underbrace{\left(1-\frac{2\mu_3}{\sum_k \mu_k}\right)}_{O(\epsilon)}\mathbf{b}'_3
-\underbrace{\frac{2\mu_2}{\sum_k \mu_k}}_{O(\epsilon)}\frac{1-\dot{s}_2}{1+\dot{s}_3} \mathbf{b}'_2
-\underbrace{\frac{2\mu_1}{\sum_k \mu_k}}_{1+O(\epsilon)}\underbrace{\frac{1-\dot{s}_1}{1+\dot{s}_3}}_{1+O(\epsilon)} \mathbf{b}'_1
\end{equation}
The zeroth order in $\epsilon$ does not vanish in $\Delta\mathbf{a}'_3$, because $\mathbf{b}'_1$ undergoes a jump. Instead we have, in generic configurations where the other coefficients are of order $1$,
$\Delta\mathbf{a}'_3\approx\Delta\mathbf{b}'_1$ which means $C_3\approx1$.

The mean values $\langle C_j \rangle$ for the set of tensions of Fig \ref{mu2light} are smaller than $1$;
\be
\langle  C_1\rangle =0.09, \qquad \langle  C_2\rangle =0.72, \qquad \langle  C_3\rangle = 0.99,
\ee
but the probability of having an amplification on strings $2$ and $3$ is significant
\be
P(C_1 > 1) =0.01 , \qquad P(C_2 > 1) = 0.19, \qquad P(C_3 > 1) = 0.47.
\ee
Again we note that, for string $3$, although amplification is frequent this is mostly limited in amplitude since the distribution is sharply peaked around a value close to $1$. 

\subsection*{Incoming kink on a heavy string with $\mu_1 \lesssim \mu_2+\mu_3$}

In the example shown in Fig  \ref{mu1heavy},  $\mu_2=\mu_3=1$ and $\mu_1=1.9$. The (superimposed) distributions $p(C_2)$, $p(C_3)$ are sharply peaked around $1$, so the amplitude of the kinks transmitted to the light strings is comparable to that of the incoming kink on the heavy string. The distribution of $C_1$ is much flatter, with a significant tail at large values. For this set of tensions one has
\be
P(C_1 > 1) = 0.11.
\ee

The presence and position of the peak in $p(C_2)$ 
can again be explained using an analytic argument. Let $\epsilon=\frac{\nu_1}{\mu}\ll 1$.  For most configurations, one has $\dot{s}_1^\pm=-1+O(\epsilon)$, while $\dot{s}_2^\pm=1-O(\epsilon)$ and $\dot{s}_3^\pm=1-O(\epsilon)$ \cite{Copeland:2006if} (see also  \eqref{spoint}). Thus from eq.~\eqref{a2},
\begin{equation}
\mathbf{a}^{\prime \pm}_2=\underbrace{\frac{1-\dot{s}_2^\pm}{1+\dot{s}_2^\pm}}_{O(\epsilon)}\left(\frac{\nu_2}{\mu}\right)\mathbf{b}'_2
-\frac{2\mu_3}{\mu}\underbrace{\frac{1-\dot{s}_3^\pm}{1+\dot{s}_2^\pm}}_{O(\epsilon)} \mathbf{b}'_3
-\underbrace{\frac{2\mu_1}{\mu}}_{1+O(\epsilon)}\underbrace{\frac{1-\dot{s}_1^\pm}{1+\dot{s}_2^\pm}}_{1+O(\epsilon)}\mathbf{b}^{\prime \pm}_1
\end{equation}
Hence to zeroth order in $\epsilon$, $\Delta\mathbf{a}^{\prime}_2\approx\Delta\mathbf{b}^{\prime}_1$ and therefore $C_2\approx1$. A similar argument applies to $C_3$.

\hspace{-.5cm}
\begin{minipage}[t]{0.48\textwidth}
 \begin{figure}[H]
   \centering
   \includegraphics[scale=0.65]{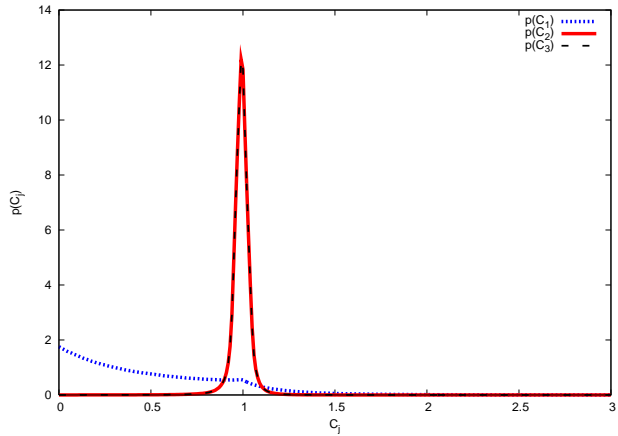} 
   \caption{Distributions of the different transmission coefficients for tensions $\mathrm{\mu_1=1.9}$ and $\mathrm{\mu_2=1=\mu_3}$. The mean values are $\langle C_1\rangle =0.49$, $\langle C_2\rangle =0.99$, $\langle C_3\rangle =0.99$ and $P(C_1<1)=0.88$, $P(C_2<1)=0.53$, $P(C_3<1)=0.53$}
  \label{mu1heavy}
  \end{figure}
\end{minipage}
\hspace{.5cm}
\begin{minipage}[t]{0.48\textwidth}
 \begin{figure}[H]
   \centering
   \includegraphics[scale=0.65]{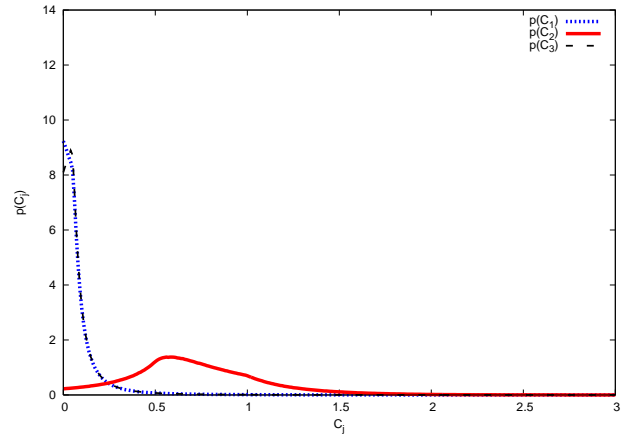} 
   \caption{The same distributions for tensions $\mathrm{\mu_1=1=\mu_3}$ and $\mathrm{\mu_2=1.9}$. Here $\langle C_1\rangle =0.09$, $\langle C_2\rangle =0.72$, $\langle C_3\rangle =0.99$ and $P(C_1<1)=0.99$, $P(C_2<1)=0.81$, $P(C_3<1)=0.99$}
  \label{mu2heavy}
  \end{figure}
\end{minipage}
\\

\subsection*{Incoming kink on light string of tension $\mu_1 \; \gsim \; \mu_2-\mu_3$}

Finally in Fig  \ref{mu2heavy},  we show the distributions of the transmission coefficients for the following set of tensions, $\mu_1=\mu_3=1$ and $\mu_2=1.9$. One sees $p(C_1)$ and $p(C_3)$ are peaked at small values. The typical amplitude of the kinks reflected on the light string $\mu_1$ and transmitted on the other light string of tebnsion $\mu_3$ is very small. The distribution of $C_2$ is flatter, so the amplitude of the kink transmitted to the heavy string depends strongly on the configuration. This distribution also has a tail for $C_2>1$, so kinks are amplified in a substantial volume of configuration space:
\be
P(C_2>1)=0.19.
\ee

Again the presence and the position of the peak in $p(C_1)$ and $p(C_3)$ can be explained analytically. The argument is as above though now $\epsilon=\frac{\nu_2}{\mu}\ll 1$ and in most configurations, $\dot{s}_1=1-O(\epsilon)$ and $\dot{s}_3=1-O(\epsilon)$ while $\dot{s}_2=-1+O(\epsilon)$.

\section*{Appendix B}
\label{ssec:tree}

Here we illustrate with a very simple toy model that an exponential behaviour of the kind exhibited in Sections \ref{section:idealised} and \ref{section:realistic} is actually very generic. 

 Consider the situation in which one constructs a tree, in which each node has two daughter nodes so that the $n$th generation therefore contains $2^n$ nodes. Each node contains an amplitude and the amplitude of each of its daughter nodes is obtained by multiplying this amplitude by a factor drawn randomly using a Bernoulli distribution: $2$ with probability $p$ or $1/2$ with probability $1-p$.  This means that there is either an amplification (by a factor of $2$) or a reduction of amplitude (by a factor of $1/2$). Since we are interested in the case where amplifications are the least probable outcome, we set
\begin{equation}
p<1/2.
\end{equation}
In order to initialize the experiment, we need to define the amplitude of the initial node (0-th generation): we set it to be $1$.

 \begin{figure}[H]
   \centering
   \includegraphics[scale=.3]{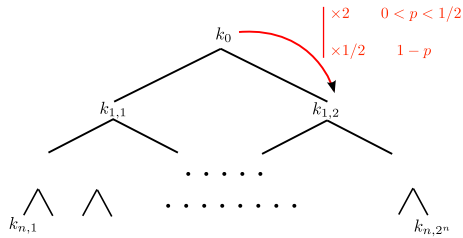} 
   \caption{Random experiment described in this section}
  \label{expevoltree}
  \end{figure}

Let $p_n$ be the probability that any given node of the $n$th generation (say $k_{1,n}$ for instance) has an amplitude larger than $1$. Since $p<1/2$, we expect $p_n$ to be small when $n$ becomes large. Our goal here is to compute analytically (in the large $n$ limit) $p_n$.

In order to be larger than $1$, $k_{1,n}$ must be the result of more amplifications than reductions so we clearly have
\begin{equation}
p_n=\sum _{k\geq n/2} \binom{n}{k} p^k (1-p)^{n-k}.
\end{equation}
Using Stirling's formula and transforming the sum into an integral using $x=k/n$, we can easily obtain
\begin{equation}
p_n=\sqrt{n}\int_{1/2}^1 \frac{1}{\sqrt{x(1-x)}}e^{n f_p(x)}dx
\end{equation}
with
\begin{equation}
f_p(x)=-x \ln(\frac{x}{p}) - (1-x)\ln(\frac{1-x}{1-p}).
\end{equation}
In the interval $[1/2,1]$, $f_p(x)$ is a decreasing function (because $p<1/2$) and therefore, it is maximal for $x=1/2$. Because of the exponential in the integral, we then expect the dominant contribution to $p_n$ to come from the vicinity of $1/2$. More precisely, by computing carefully the integral, it is possible to show that 
\begin{equation}
p_n \underset{n\rightarrow +\infty}{\sim} \frac{2}{-f_p'(1/2)}\frac{\sqrt{n}}{n} e^{n f_p(1/2)} = \frac{2}{\ln(\frac{1-p}{p})}\frac{1}{\sqrt{n}} \left(2\sqrt{p(1-p)}\right)^n .
\end{equation}

As expected, $p_n$ decreases exponentially since $2\sqrt{p(1-p)}<1$. However, we are interested in the \emph{number} $N_n$ of nodes of the $n$-th generation that have an amplitude larger than $1$. If $n$ is large enough, this number will typically be $2^n p_n$:
\begin{equation}
N_n \underset{n\rightarrow +\infty}{\sim}\frac{2}{\ln(\frac{1-p}{p})}\frac{1}{\sqrt{n}} \left(4\sqrt{p(1-p)}\right)^n .
\end{equation}
This number \emph{increases exponentially} when $4\sqrt{p(1-p)}>1$ i.e. $p(1-p)>\frac{1}{16}$. This condition is satisfied as soon as $p>\frac{1}{2}-\frac{\sqrt{3}}{4}\approx 0.07$ (remember that we also imposed $p<1/2$).

This means that even if amplifications are not the most probable outcome of amplitude transmissions (as is the case for our physical system), the number of nodes of the $n$th generation that have an amplitude larger than some fixed value increases exponentially provided that $p$ is not too small, i.e. provided that amplifications are not too rare.

Note that the fraction of nodes that have an amplitude larger than $1$ is given by $p_n$ and tends to $0$ exponentially fast and that the proliferation of large amplitude nodes is only possible because the total number ($2^n$ in our example) increases faster than $p_n$ decreases. As expected, the very large majority of nodes have small amplitude.

Of course, this very simple example is not a good physical picture of our kink proliferation (amplitudes can be larger than 1 in this example). Indeed, the Bernoulli distribution is a crude simplification of $p_{A_i}(C_1,C_2,C_3)$ which does not take into account the fact that the amplitude of the daughter kinks are actually correlated and forgets about the amplitude of the incoming kink $A_i$. However, we expect the general mechanism to remain the same.


\begin{thebibliography}{}



\bibitem{Polchinski:2004hb}
J.~Polchinski,
%``Cosmic superstrings revisited,''
AIP Conf.\ Proc.\  {\bf 743}, 331 (2005)
[Int.\ J.\ Mod.\ Phys.\ A {\bf 20}, 3413 (2005)]
[arXiv:hep-th/0410082].

\bibitem{Kibble:2004hq}
T.~W.~B.~Kibble,
%``Cosmic strings reborn?,''
Lecture at COSLAB 2004, Ambleside, 10--17 September 2004 (unpublished),
arXiv:astro-ph/0410073.

\bibitem{Davis:2005dd}
A.~C.~Davis and T.~W.~B.~Kibble,
%``Fundamental cosmic strings,''
Contemp.\ Phys.\  {\bf 46}, 313 (2005)
[arXiv:hep-th/0505050].
%

%\cite{Myers:1900zz}
\bibitem{Myers:1900zz}
  R.~C.~Myers and M.~Wyman,
  %``Cosmic superstrings,''
%\href{/spires/find/hep/www?irn=8453853}{SPIRES entry}
{\it  In *Erdmenger, J. (ed.): String cosmology* 121-156}

%\cite{Copeland:2009ga}
\bibitem{Copeland:2009ga}
  E.~J.~Copeland and T.~W.~B.~Kibble,
  %``Cosmic Strings and Superstrings,''
  Proc.\ Roy.\ Soc.\ Lond.\  A {\bf 466} (2010) 623
  [arXiv:0911.1345 [hep-th]].
  %%CITATION = PRSLA,A466,623;%%


\bibitem{Tye:2005fn}
  S.~H.~Tye, I.~Wasserman and M.~Wyman,
  %``Scaling of multi-tension cosmic superstring networks,''
  Phys.\ Rev.\  D {\bf 71}, 103508 (2005)
  [Erratum-ibid.\  D {\bf 71}, 129906 (2005)]
  [arXiv:astro-ph/0503506].
  %%CITATION = PHRVA,D71,103508;%%
  
  \bibitem{Hindmarsh:2006qn}
  M.~Hindmarsh and P.~M.~Saffin,
  %``Scaling in a SU(2)/Z(3) model of cosmic superstring networks,''
  JHEP {\bf 0608} (2006) 066
  [arXiv:hep-th/0605014].
  %%CITATION = JHEPA,0608,066;%%
  
  
  \bibitem{Copeland:2005cy}
  E.~J.~Copeland and P.~M.~Saffin,
  %``On the evolution of cosmic-superstring networks,''
  JHEP {\bf 0511} (2005) 023
  [arXiv:hep-th/0505110].
  %%CITATION = JHEPA,0511,023;%%
  
  \bibitem{Urrestilla:2007yw}
  J.~Urrestilla and A.~Vilenkin,
  %``Evolution of cosmic superstring networks: a numerical simulation,''
  JHEP {\bf 0802} (2008) 037
  [arXiv:0712.1146 [hep-th]].
  %%CITATION = JHEPA,0802,037;%%

\bibitem{Avgoustidis:2007aa}
  A.~Avgoustidis and E.~P.~S.~Shellard,
  %``Velocity-Dependent Models for Non-Abelian/Entangled String Networks,''
  Phys.\ Rev.\  D {\bf 78}, 103510 (2008)
  [Erratum-ibid.\  D {\bf 80}, 129907 (2009)]
  [arXiv:0705.3395 [astro-ph]].
  %%CITATION = PHRVA,D78,103510;%%


\bibitem{Avgoustidis:2009ke}
  A.~Avgoustidis and E.~J.~Copeland,
  %``The effect of kinematic constraints on multi-tension string network
  %evolution,''
  Phys.\ Rev.\  D {\bf 81}, 063517 (2010)
  [arXiv:0912.4004 [hep-ph]].
  %%CITATION = PHRVA,D81,063517;%%



\bibitem{DV1}
  T.~Damour and A.~Vilenkin,
  %``Gravitational radiation from cosmic (super)strings: Bursts, stochastic
  %background, and observational windows,''
  Phys.\ Rev.\  D {\bf 71} (2005) 063510
  [arXiv:hep-th/0410222].
  %%CITATION = PHRVA,D71,063510;%%

%\cite{Siemens:2006vk}
\bibitem{Siemens:2006vk}
  X.~Siemens, J.~Creighton, I.~Maor, S.~Ray Majumder, K.~Cannon and J.~Read,
  %``Gravitational wave bursts from cosmic (super)strings: Quantitative
  %analysis and constraints,''
  Phys.\ Rev.\  D {\bf 73} (2006) 105001
  [arXiv:gr-qc/0603115].
  %%CITATION = PHRVA,D73,105001;%%


%\cite{Siemens:2006yp}
\bibitem{Siemens:2006yp}
  X.~Siemens, V.~Mandic and J.~Creighton,
  %``Gravitational wave stochastic background from cosmic (super)strings,''
  Phys.\ Rev.\ Lett.\  {\bf 98} (2007) 111101
  [arXiv:astro-ph/0610920].
  %%CITATION = PRLTA,98,111101;%% 

\bibitem{Jackson}
M.~G.~Jackson and X.~Siemens,
  %``Gravitational Wave Bursts from Cosmic Superstring Reconnections,''
  JHEP {\bf 0906} (2009) 089
  [arXiv:0901.0867 [hep-th]].
  %%CITATION = JHEPA,0906,089;%%
%\cite{Jackson:2004zg}


\bibitem{Jackson:2004zg}
  M.~G.~Jackson, N.~T.~Jones and J.~Polchinski,
  %``Collisions of cosmic F- and D-strings,''
  JHEP {\bf 0510} (2005) 013
  [arXiv:hep-th/0405229].
  %%CITATION = JHEPA,0510,013;%%




\bibitem{CKS}
  E.~J.~Copeland, T.~W.~B.~Kibble and D.~A.~Steer,
  %``Collisions of strings with Y junctions,''
  Phys.\ Rev.\ Lett.\  {\bf 97} (2006) 021602
  [arXiv:hep-th/0601153].
  %%CITATION = PRLTA,97,021602;%%
  
  
  
   
  
  \bibitem{CKS2}
  E.~J.~Copeland, H.~Firouzjahi, T.~W.~B.~Kibble and D.~A.~Steer,
  %``On the Collision of Cosmic Superstrings,''
  Phys.\ Rev.\  D {\bf 77} (2008) 063521
  [arXiv:0712.0808 [hep-th]].
  %%CITATION = PHRVA,D77,063521;%%
    
  
\bibitem{Binetruy10b}
P.~Binetruy, A.~Bohe, T.~Hertog and D.~A.~Steer,
%``Gravity Wave Signatures of Kink Proliferation on Cosmic Strings with Junctions,''
in progress

 \bibitem{DePies:2007bm}
   M.~DePies, and C.~Hogan,
  %   title     = "{Stochastic Gravitational Wave Background from Light Cosmic
 %                 Strings}",
  Phys.\ Rev.\  D {\bf 75} (2007) 125006
  [arXiv:astro-ph/0702335].

  \bibitem{Damour}
  T.~Damour and A.~Vilenkin,
  %``Gravitational wave bursts from cusps and kinks on cosmic strings,''
  Phys.\ Rev.\  D {\bf 64} (2001) 064008
  [arXiv:gr-qc/0104026].
  %%CITATION = PHRVA,D64,064008;%%
   
  
\bibitem{Us}
P.~Binetruy, A.~Bohe, T.~Hertog and D.~A.~Steer,
  %``Gravitational Wave Bursts from Cosmic Superstrings with Y-junctions,''
  Phys.\ Rev.\  D {\bf 80} (2009) 123510
  [arXiv:0907.4522 [hep-th]].
  %%CITATION = PHRVA,D80,123510;%%
  
  
  \bibitem{Mairii}
  A.~C.~Davis, W.~Nelson, S.~Rajamanoharan and M.~Sakellariadou,
  %``Cusps on cosmic superstrings with junctions,''
  JCAP {\bf 0811} (2008) 022
  [arXiv:0809.2263 [hep-th]].
  %%CITATION = JCAPA,0811,022;%%
  
  %\cite{O'Callaghan:2010ww}
\bibitem{Ruth}
  E.~O'Callaghan, S.~Chadburn, G.~Geshnizjani, R.~Gregory and I.~Zavala,
  %``On detection of extra dimensions with gravity waves from cosmic strings,''
  arXiv:1003.4395 [hep-th].
  %%CITATION = ARXIV:1003.4395;%%



 \bibitem{Bevis:2009az}
  N.~Bevis, E.~J.~Copeland, P.~Y.~Martin, G.~Niz, A.~Pourtsidou, P.~M.~Saffin and D.~A.~Steer,
  %``Evolution and stability of cosmic string loops with Y-junctions,''
  Phys.\ Rev.\  D {\bf 80} (2009) 125030
  [arXiv:0904.2127 [hep-th]].
  %%CITATION = PHRVA,D80,125030;%%
  
  
  \bibitem{Copeland:2006if}
  E.~J.~Copeland, T.~W.~B.~Kibble and D.~A.~Steer,
  %``Constraints on string networks with junctions,''
  Phys.\ Rev.\  D {\bf 75} (2007) 065024
  [arXiv:hep-th/0611243].
  %%CITATION = PHRVA,D75,065024;%%
  
%%%%%%%%%%%%%%%%%%%%%%%%%%%%%%%%%%%%%%%  
  
  
\bibitem{Hassan1}
R.Brandenberger, H. Firouzajhi, J.Karouby and S.Khosravi, 	
%``Zipping and Unzipping of Cosmic String Loops in Collision'',
Phys.\ Rev.\ D{\bf 80} (2009) 083508
[arXiv:0907.4986v2 [hep-th]].





  
  
\bibitem{Copeland:2009dk}
  E.~J.~Copeland and T.~W.~B.~Kibble,
 % ``Kinks and small-scale structure on cosmic strings,''
  Phys.\ Rev.\  D {\bf 80} (2009) 123523
  [arXiv:0909.1960 [astro-ph.CO]].
  %%CITATION = PHRVA,D80,123523;%%  



%%%%%%GW emission in Cosmic string and cosmic superstring networks%%%%%%%%

%%%from reconnections%%%%%
%
%\bibitem{MarkJ}
%M.G.~Jackson and X~Siemens, 
%``Gravitational wave bursts from cosmic superstring
%reconnections'', [arXiv:0901.0867[hep-th]].


  
  
  
  
  
 
   
  \bibitem{KT82} T.W.B. Kibble and N. Turok, Phys. Lett. 116B (1982) 141.

\bibitem{Tur84} N. Turok, Nucl. Phys. B 242 (1984) 520.

\bibitem{CDH} A. L. Chen, D. A. DiCarlo and S. A. Hotes, Phys. Rev. D 37
(1988) 863.

\bibitem{DES} D. DeLaney, K. Engle and X. Scheick, Phys. Rev.
D 41 (1990) 1775.

  
  
\bibitem{Siemens:1994ir}
  X.~A.~Siemens and T.~W.~B.~Kibble,
  %``High harmonic configurations of cosmic strings: An Analysis of
  %selfintersections,''
  Nucl.\ Phys.\  B {\bf 438} (1995) 307
  [arXiv:hep-ph/9412216].
  %%CITATION = NUPHA,B438,307;%% 

  


\end{thebibliography}
\end{document}